\documentclass{article}
\usepackage[utf8]{inputenc}
\usepackage[english]{babel}
\usepackage{csquotes}

\usepackage{amsmath,amssymb,xcolor}
\usepackage{enumitem}

\usepackage[backend=biber,style=alphabetic,maxnames=10]{biblatex}
\AtBeginBibliography{\setlength{\emergencystretch}{2em}}
\usepackage{amsthm,thmtools}
\declaretheoremstyle[qed=$\lrcorner$,bodyfont=\it]{it}
\declaretheoremstyle[qed=$\lrcorner$,bodyfont=\rm]{rm}

\declaretheorem[style=it,numberwithin=subsection]{lemma}

\makeatletter
\let\c@equation=\c@lemma

\let\c@subsubsection=\c@lemma

\makeatother

\declaretheorem[style=it,numberlike=lemma]{corollary}
\declaretheorem[style=it,numberlike=lemma]{theorem}
\declaretheorem[style=it,numberlike=lemma]{proposition}
\declaretheorem[style=it,numberlike=lemma]{remark}

\declaretheorem[style=it,name=Theorem]{theoremA}

\usepackage[hypertexnames=false, % to make ids unique and avoid warnings
hyperfootnotes=false,
colorlinks,
linkcolor={blue},
citecolor={blue},
urlcolor={blue}
]{hyperref}
\usepackage[hypertexnames=false]{hyperref}
\input{defs}
\mkmathletters{cal}{\Lletters}   %\Acal,\bcal caligraphic
\mkmathletters{bf}{\Lletters}   %\Abf,\bbf   bold
\mkmathletters{bb}{\Lletters}   %\Abb,\Zbb   blackboard bold
\mkmathletters{rm}{\Lletters}   %\Arm, \brm  roman
\mkmathletters{sf}{\Lletters}   %\Asf, \bsf  san serif
\mkmathletters{bar}{\Lletters}  % with bar,  smaller then overlined

\dmo\aut{\mathrm{Aut}}
\dmo\Hom{\mathrm{Hom}}
\dmo\ing{\mathsf{ing}}
\dmo\supp{\mathrm{supp}}
\dmo\ext{\mathrm{Ext}}
\dmo\convexhull{\mathrm{ConvexHull}}

\def\1{\bm{1}}

\begin{document}
\title{Shapes and Norms of Random Pairs}
\author{Rostislav Matveev}

\maketitle

\begin{abstract}
  The shape function of a pair of finite-valued random variables was
  introduced in~\cite{matveev2026beyond}, where it was used to derive
  a spectral bound on the entanglement of the pair, a quantity
  measuring the extent to which their mutual information can be
  extracted. In this article, we further develop the theory of shape
  functions for pairs of random variables. We prove that, when $(X,Y)$
  is uniformly supported on the edges of a biregular bipartite graph,
  the value of the shape function $\Scal(X,Y)(\alpha,\beta)$ equals
  the logarithm of the operator norm of the graph's incidence matrix
  with respect to Lebesgue exponents determined by
  $(\alpha,\beta)$.This identification, in particular, enables the
  numerical approximation of the shape function and, by duality, of
  the extension profile, also known as the tension region, of the
  pair.  We also establish a collection of relations and inequalities
  satisfied by shape functions, including convexity and monotonicity
  properties, composition inequalities, and relations describing their
  behavior under conditioning and the adjoining of variables.
\end{abstract}

% \tableofcontents
\section{Shape function and its relation to the norms of incidence operators}
\label{s:intro}
The \emph{shape function} of a pair of jointly distributed random
variables $(X,Y)$, introduced in~\cite{matveev2026beyond}, is a
function on $[0,1]^{2}$ defined by
\[
  \Scal(X,Y)(\alpha,\beta)
  :=
  \sup\set{\alpha\cdot H(X|W)+\beta\cdot H(Y|W)-I(X:Y|W)}
\]
where supremum is taken over all extensions $(X,Y,W)$.
Function $\Scal(X,Y)$ is convex on the square and it is affine on the
upper-right triangle
\[
  \set{(\alpha,\beta)\st \alpha+\beta\geq 1}\cap[0,1]^{2}
\]
The affine part is determined by the entropy profile of the pair, that is
entropies of the two variables and their joint. In general, the
restriction of $\Scal(X,Y)$ to the lower-left triangle is not
determined by the entropy profile and contains additional information
about the pair.  In particular, the \emph{entanglement} of the pair
$(X,Y)$ -- the quantity that measures to what extend mutual
information is extractable and defined by
\begin{equation*}
  \Ecal(X,Y)=\sup_{W}\set{I(X:Y|W)+I(W:Y|X)+I(X:W|Y)}
\end{equation*}
can be recovered from $\Scal(X,Y)$, see~\cite{matveev2026beyond,
  matveev2026spectral} and Equation~\eqref{eq:mmrv} on
page~\pageref{eq:mmrv} of the present article.
 
If the pair $(X,Y)$ is uniform on its support, that is, each of the
random variables $X$, $Y$ and their joint $XY$ are uniform on their
respective supports,
then it is completely determined by the supporting graph
$\Gsf=(\Xsf\sqcup\Ysf,\Esf)$, where
\[
  \Xsf:=\supp X,
  \qquad
  \Ysf:=\supp Y,
  \qquad
  \Esf:=\supp XY
\]
Such a bipartite graph, which is necessarily biregular, can be
represented by its bipartite incidence matrix $M$ of size
$|\Xsf|\times|\Ysf|$.

In this article we evaluate the shape function of a pair uniform on
its support in terms of the incidence matrix of the supporting
graph. We view $M$ as a matrix, with respect to standard
bases, of the bilinear form
\begin{equation*}
  M:\ell^{p}_\Xsf\times\ell^{q}_\Ysf\to\Rbb,\qquad
  \phi M\psi:=\sum_{(\xsf,\ysf)\in\Esf}\phi(\xsf)\psi(\ysf) 
\end{equation*}
where $p,q\in[1,\infty]$ and where we use infix notation $\phi M\psi$
for evaluation of the form. We define its $(p,q)$-norm by
\begin{equation*}
  \|M\|_{p,q}
  :=
  \sup\set{\frac{|\phi M\psi|}{\|\phi\|_{p}\cdot\|\psi\|_{q}}\st
    0\neq\phi\in\ell^{p}_\Xsf,\;0\neq\psi\in\ell^{q}_\Ysf}
\end{equation*}

The main result of this article is the following theorem.
\begin{theoremA}\label{p:intro-main}
  Let $(X,Y)$ be a pair of random variables uniformly supported on a
  homogeneous bipartite graph $\Gsf=(\Xsf\sqcup\Ysf,\Esf)$ with
  the incidence form $M$. Then for all $\alpha,\beta\in
  [0,1]$ holds
  \[
    \Scal(X,Y)(\alpha,\beta)=\log\|M\|_{p,q}
  \]
  where $1/p=1-\alpha$ and $1/q=1-\beta$.
\end{theoremA}

By Tropical Asymptotic Equipartition
Property,~\cite{matveev2018asymptotic}, a pair $(X^{n},Y^{n})$
obtained by taking $n$ independent copies of a pair $(X,Y)$, can be
arbitrarily well approximated on the normalized scale by a pair of
random variables uniformly supported on a homogeneous graph. Since the
shape is stable, that is,
\[
  \Scal(X^{n},Y^{n})=n\cdot\Scal(X,Y),
\]
and continuous with respect to such approximations,
Theorem~\ref{p:intro-main} allows one, in principle, to evaluate shape
function for an arbitrary, not necessarily uniformly supported, pair.

Theorem~\ref{p:intro-main} also gives a practical way to approximate
the shape function. Indeed, the norm of the bilinear form
$\|M\|_{p,q}$ is the same as the operator norm $\|M\|_{p\to q'}$,
where $q'$ is the Hölder conjugate of $q$.  Lebesgue operator norms
$\|M\|_{p\to q'}$ can be approximated numerically by robust methods.
Thus, for pairs uniformly supported on biregular bipartite graphs, one
can numerically approximate both the shape function and, by duality,
the extension profile.

In the next section we introduce our notation and
conventions. Section~\ref{s:norms-tensor} we recall some facts about
norms on tensor product of normed vector spaces and prove
multiplicativity of Lebesgue operator norms in the hypercontractive
regime, Proposition~\ref{p:multiplicative}. Further we prove our main
technical tool, Theorem~\ref{p:almost-resonance-pairs}, that asserts
that for high tensor powers of bilinear forms there are uniform
vectors which are almost in resonance. In
Section~\ref{s:graph-bilinear} we apply the results of the previous
sections to bilinear forms associated to bipartite graphs.
Theorem~\ref{p:intro-main} is proven in Section~\ref{s:main-thm}. 
We also derive inequalities satisfied by the shape function in
Section~\ref{s:properties}. Especially interesting is the
inequality~\eqref{eq:comp} on page~\pageref{eq:comp}. We do not know
whether this inequality can be derived in the purely entropic context,
that is without using Theorem~\ref{p:intro-main}.

\section{Notation, conventions, recollections}
\label{s:prelim}
\subsection{Notation}
\label{s:notation}
For a natural number $n$ we denote $[n]:=\set{0,\dots,n-1}$.  We write
$\sharp S$ or $|S|$ for the cardinality of a finite set $S$.  Denote
the $\log$-cardinality of a set by $[S]:=\log|S|$.  For a finite set
$S$, we write $\Rbb^S$ for the vector space of real-valued functions
on $S$. For a subset $T\subset S$ and a point $s\in S$ we denote by
$\1_{T}\in\Rbb^{S}$ the indicator function of $T$ and by
$\delta_{s}:=\1_{\set{s}}$ the point mass.

For an extended real $p\in[1,\infty]$ we denote by $p'$ its Hölder
conjugate, that is the two extended reals $p,p'$ must satisfy
$1/p+1/p'=1$.

We use logarithms with the natural base throughout the article
and $\ebf$ denotes Euler's number.

\subsection{Random variables}
\label{s:rv}
All random variables in this article have finite alphabets. As a
notational convention, use capitals $X,Y,\dots$ for random variables
and the corresponding san-serif letters $\Xsf,\Ysf,\dots$ for their
alphabets.  For a random variable $X$ and $\xsf\in\Xsf$ we use
$p_X(\xsf):=P[X=\xsf]$.  We call $\xsf\in\Xsf$ an \emph{atom} of $X$
if $p_{X}(\xsf)>0$.The \emph{support} $\supp X$ is the collection of
all atoms.

% We denote by $\sharp X$ the cardinality of the support of random
% variable $X$, thus $\sharp X:=\sharp\supp X$.

We tacitly assume that the alphabets of different random variables are
disjoint and we write $(X|\ysf)$ for the conditional random variable
in lieu of $(X|Y=\ysf)$.

Tuples of random variables are written as comma-separated lists, such
as, for example, $(X,Y,Z)$ or $(X_{i}\st i\in[n])$, while joints of random
variables, regarded as a single random variable, where marginalization
structures are ignored, are denoted by concatenation of the
corresponding letters, such as $XY$ or $XY\!Z$, or by using the subset for
the subscript, as in $X_{I}$ for the joint of
$(X_{i}:i\in I)$, $I\subset[n]$.  For example, for a \emph{triple} of
random variables, $(X,Y,Z)$, the notation $(X,Y\!Z)$ stands for a
\emph{pair} of variables consisting of variable $X$ and the joint
variable $Y\!Z$. This pair is \emph{different} from the \emph{pair}
$(XY,Z)$. Given a pair $(X,Y)$, a third random variable $W$ jointly
distributed with $(X,Y)$ forms an \emph{extension} $(X,Y,W)$ of the
pair and is called an \emph{extending} variable.

% \red{Do I need this?}
% Jointly distributed random variables $(X_{i}\st i\in[n])$ satisfy a
% series of so called Shannon inequalities
% \begin{equation*}
  % I(X_{I}:X_{J}|X_{K})\geq 0,
  % \qquad
  % I,J,K\subset[n]
% \end{equation*}
% where cases of empty or coinciding $I,J,K$ are allowed.  A, not
% necessarily linear, inequality on entropies of joints, $H(X_I)$,
% $I\subset[k]$ is called \emph{Shannon-type} if it is satified by rank
% functions of all polymatroids. All Shannon-type
% inequalities used in the article can be verified either by hand or by using a
% general purpose linear programming kit, or by using software packages
% especially designed for the purpose, such as~\cite{minitip, PSITIP,
  % pulikkoonattu2008xitip,pulikkoonattu2020aitip,yeung1996itip}. A
% detailed discussion of these and other software packages can be found
% in~\cite{yeung2021machine}.

Supports of the variables in the pair $(X,Y)$ form a bipartite graph
$\Gsf=(\Xsf\sqcup\Ysf,\Esf)$, where
\begin{equation*}
  \Xsf:=\supp X,
  \qquad
  \Ysf:=\supp Y,
  \qquad
  \Esf:=\supp XY
\end{equation*}
We say that $(X,Y)$ is \emph{supported} on $\Gsf$ and write
\begin{equation*}
  \Gsf = \supp(X,Y)
\end{equation*}

A tuple of random variables $(X_{i}\st i\in[n])$ is called
\emph{uniform on its support} if all partial joints $X_{I}$,
$I\subset[n]$ are uniform on their respective supports.  If the pair
$(X,Y)$ is uniform on its support, then the supporting graph
$\Gsf=\supp(X,Y)$ is biregular and its combinatorial structure
completely determines the pair. In that case we say that $(X,Y)$
\emph{uniformly supported on $\Gsf$}.

% \red{[Used?]}We say that a random variable $X$ is \emph{$\delta$-uniform} for
% $\delta\geq1$ if the probabilities of any two atoms of $X$ differ by at
% most by a factor of $\delta$:
% \begin{equation*}
  % \frac{\max_{\xsf} p_{X}(\xsf)}{\min_{\xsf}p_{X}(\xsf)}\leq\delta
% \end{equation*}
% where extrema are taken over the support of $X$.
% If $X$ is $\delta$-uniform, then
% \begin{equation*}
  % \log\sharp X -\log\delta\leq H(X)\leq \log\sharp X
% \end{equation*}
% A tuple of random variables $(X_{i}\st i\in[n])$ is called
% \emph{$\delta$-uniform} if all joints $X_{I}$, $I\subset[n]$, are $\delta$-uniform.

\subsection{Graphs}
\label{s:graphs}
All graphs considered in this article have no isolated vertices. 
A bipartite graph $\Gsf=(\Xsf\sqcup\Ysf,\Esf)$ is called
\emph{biregular} if the degrees of vertices are constant within each
part. We denote by $d_{1}(\Gsf),d_{2}(\Gsf)$ the left and right
degrees of $\Gsf$, respectively. The \emph{automorphism group}
$\aut(\Gsf)$ of $\Gsf$ is the group of symmetries of $\Gsf$ preserving
each part. Graph is called \emph{homogeneous} if $\aut(\Gsf)$ acts
transitively on the edge set $\Esf$. Homogeneous graphs are biregular.
By a subgraph $\Hsf\subset\Gsf$ we always mean a nonempty subgraph
without isolated vertices. We denote by $\Xsf_{\Hsf}$, $\Ysf_{\Hsf}$
and $\Esf_{\Hsf}$ the left, right parts and edge-set of $\Hsf$,
respectively. We also set
\begin{align*}
  &[\Xsf_{\Hsf}]:=\log|\Xsf_{\Hsf}|, 
  &&[\Esf_{\Hsf}]:=\log|\Esf_{\Hsf}|, \\
  &[\Ysf_{\Hsf}]:=\log|\Ysf_{\Hsf}|, 
  &&[\Xsf_{\Hsf}:\Ysf_{\Hsf}]:=\log\frac{|\Xsf_{\Hsf}|\cdot|\Ysf_{\Hsf}|}{|\Esf_{\Hsf}|}
\end{align*}

For a pair $(X,Y)$ uniformly supported on a graph
$\Gsf=(\Xsf\sqcup\Ysf,\Esf)$ the following identities hold:
\begin{align*}
  &H(X)=[\Xsf],
  &&H(XY)=[\Esf],
  &&H(Y|X)=\log d_{1}(\Gsf),\\
  &H(Y)=[\Ysf],
  &&I(X:Y)=[\Xsf:\Ysf],
  &&H(X|Y)=\log d_{2}(\Gsf)   
\end{align*}

\subsection{Extension profile and shape function}
\label{s:extension-shape}
The \emph{entropy profile} of a pair $(X,Y)$ is a vector
\begin{equation*}
  e(X,Y):=\vect{H(X)\\H(Y)\\I(X:Y)}\in\Rbb^{3}
\end{equation*}
For an extension $(X,Y,W)$ the \emph{conditional entropy profile} is
\begin{equation*}
  e(X,Y|W):=\vect{H(X|W)\\H(Y|W)\\I(X:Y|W)}
\end{equation*}
The \emph{extension profile} of a pair $(X,Y)$ is the set of all
conditional entropy profiles for all extensions of the pair.
\begin{equation*}
  \ext(X,Y)
  :=
  \set{e(X,Y|W)\st \text{$(X,Y,W)$ is an extension of $(X,Y)$}}
  \subset
  \Rbb^{3}
\end{equation*}
It is closely related to the \emph{tension region};
see~\cite{prabhakaran2014assisted,li2017extended,csirmaz2023short}.
One may think of the extending variable $W$ as a probe testing finer
properties of the relation between variables $X$ and $Y$ --- those
that are not already reflected in the entropy profile of the pair.

The extension profile of any pair is a convex compact subset of
$\Rbb^{3}$.  In~\cite{matveev2026beyond} a dual object, the so called
\emph{shape function}, or simply \emph{shape} is introduced and
studied. It is the function on the square $[0,1]^{2}$ in the
$(\alpha,\beta)$-plane defined by
\begin{align*}
  \Scal(X,Y)(\alpha,\beta)
  &:=
    \sup\set{\alpha\cdot x+\beta\cdot y - z\st (x,y,z)\in\ext(X,Y)}\\
  &\phantom{:}=
    \sup_{W}\set{\alpha\cdot H(X|Y)+\beta\cdot H(Y|W) - I(X:Y|W)}
\end{align*}
where the later supremum is over all extending variables $W$.

For the basic properties of $\Scal(X,Y)$, we refer the reader
to~\cite{matveev2026beyond}. There, the shape function is studied in
detail, including its upper and lower bounds and, for pairs uniformly
supported on graphs, its relation to spectral properties of the
supporting graph. In this article we establish some additional properties
of the shape.

\section{Norms on tensor products}
\label{s:norms-tensor}
Here we recall several standard facts about tensor products of normed
vector spaces.  We restrict attention to finite-dimensional spaces,
although many of the constructions discussed below extend to general
Banach spaces with the usual additional analytic care.  Standard
references for tensor products of Banach spaces
include~\cite{defant1993tensor, ryan2002introduction}.

The main results of this section are
Theorem~\ref{p:almost-resonance-pairs} and
Corollary~\ref{p:unif-dense}. The main technical tool,
Proposition~\ref{p:multiplicative}, is, most likely, known to the
specialists, but for the lack of a suitable reference we provide a
proof.

\subsection{Tensor products}
\label{s:tensor-product}
We write $A\iso[a]B$ for an algebraic isomorphism of vector spaces,
and $A\iso[b]B$ for an isomorphism of Banach spaces.

We use several natural algebraic isomorphisms between tensor products
of finite-dimensional vector spaces.  Since these isomorphisms are
canonical, we shall use them implicitly and suppress them from the
notation:
\begin{align}
  \tag{commutativity}
    A\otimes B&\iso[a] B\otimes A,\\
  \tag{associativity}
    (A\otimes B)\otimes C&\iso[a] A\otimes (B\otimes C),\\
  \tag{$\text{duality}_1$}
    (A\otimes B)^{*}&\iso[a] A^{*}\otimes B^{*},\\
  \tag{$\text{duality}_{2}$}
    \Hom(A,B)&\iso[a] A^{*}\otimes B .
\end{align}
For example, for our purposes a linear map $M:A\to B$, its adjoint
$M^{*}\colon B^{*}\to A^{*}$, the bilinear form
\[
  B^{*}\times A\to\Rbb,
  \qquad
  (b^{*},a)\mapsto b^{*}Ma,
\]
and the tensor in $A^{*}\otimes B$ are regarded as different
realizations of the same object, denoted by $M$.

We shall also use the following algebraic identifications for spaces of
functions on finite sets:
\begin{equation*}
  \Rbb^{\Xsf}\otimes\Rbb^{\Ysf}
  \iso[a]
  \Rbb^{\Xsf\times\Ysf}
  \iso[a]
  \left(\Rbb^{\Xsf}\right)^{\!\Ysf}
\end{equation*}

\subsection{Cross norms}
\label{s:cross-norms}
\subsubsection{Algebraic definition of cross norms}
\label{s:cross-alg}
Let $(A,\alpha)$ and $(B,\beta)$ be Banach spaces. In this article, we
call a norm $\gamma$ on the algebraic tensor product $A\otimes B$ a
\emph{cross norm}%
\footnote{Sometimes in the literature a \emph{cross norm} is defined
  as a norm on the tensor product satisfying only the first condition
  in~\eqref{eq:crossnorm}, while a norm satisfying both conditions is
  called a \emph{reasonable cross norm}.} %
if, for all $a\in A$, $b\in B$, $a^{*}\in A^{*}$, and
$b^{*}\in B^{*}$, one has
\begin{equation}
  \label{eq:crossnorm}
  \begin{aligned}
    \gamma(a\otimes b)
    &= \alpha(a)\cdot\beta(b),\\
    \gamma^{*}(a^{*}\otimes b^{*})
    &= \alpha^{*}(a^{*})\cdot\beta^{*}(b^{*})
  \end{aligned}
\end{equation}
where $\alpha^*$ stands for the dual norm on $A^*$, etc.

\subsubsection{Geometric definition of cross norms}
\label{s:cross-geo}

For the geometrically oriented reader, cross norms admit the following
equivalent description. For a Banach space $(A,\alpha)$, denote by
$\Sbb_{\alpha}\subset A$ the $\alpha$-unit sphere.  The projectivized
product
\[
  P(\Sbb_{\alpha}\times\Sbb_{\beta})
  :=
  \set{a\otimes b:\ a\in\Sbb_{\alpha},\ b\in\Sbb_{\beta}}
\]
is naturally contained in the tensor product $A\otimes B$.

A norm $\gamma$ on $A\otimes B$ is a \emph{cross norm} if and only if
\begin{equation*}
  P(\Sbb_{\alpha}\times\Sbb_{\beta})
  \subset \Sbb_{\gamma}
  \qquad\text{and}\qquad
  P(\Sbb_{\alpha^{*}}\times\Sbb_{\beta^{*}})
    \subset \Sbb_{\gamma^{*}}
\end{equation*}

\subsubsection{Injective and projective cross norms}
\label{s:inj-proj}
Here we consider two notable examples of cross norms: the
least/injective and the greatest/projec\-tive cross norms.%
\footnote{These two norms are often denoted by $\epsilon$ and $\pi$,
  respectively. We use different notation to emphasize the dependence
  of the injective and projective tensor norms on $\alpha$ and
  $\beta$.}
\begin{align*}
  \tag{injective}
  \alpha\vee\beta(v)
  &:=
    \sup\set{(a^{*}\otimes b^{*})(v)\st
    a^{*}\in\Bbb_{\alpha^{*}},\;b^{*}\in\Bbb_{\beta^{*}}}\\
  \tag{projective}
  \alpha\wedge\beta(v)
  &:=
    \inf\set{\sum\alpha(a_{i})\beta(b_{i})\st
    v=\sum a_{i}\otimes b_{i}}
\end{align*}
where $v\in A\otimes B$.

The operations $\vee$ and $\wedge$ are dual to each other in the sense
that
\begin{equation*}
  (\alpha\vee\beta)^{*}=\alpha^{*}\wedge\beta^{*}
  \qquad\text{and}\qquad
  (\alpha\wedge\beta)^{*}=\alpha^{*}\vee\beta^{*}
\end{equation*}
The projective norm can be geometrically defined by the following. We
declare the $\alpha\wedge\beta$-unit ball to be the smallest convex set
containing the projectivized product of the unit balls in $A$ and $B$,
that is
\begin{equation*}
  \Bbb_{\alpha\wedge\beta}:=\convexhull P(\Bbb_{\alpha}\times\Bbb_{\beta})
\end{equation*}
where $\Bbb_{\alpha}$ stands for the closed unit ball in a Banach
space $(A,\alpha)$, etc.
By duality there is a similar description of the injective norm
$\alpha\vee\beta$.

Note that for an operator $M:(A,\alpha)\to(B,\beta)$ (which can be
considered as an element of $A^{*}\otimes B$) the operator norm is
\[
  \|M\|_{\alpha\to\beta}=(\alpha^{*}\vee\beta)(M)
\]

It is known and follows directly from the geometric description
of the projective and injective norms that a norm
$\gamma$ on $A\otimes B$ is a cross norm if and only if
\begin{equation*}
  \alpha\vee\beta\leq\gamma\leq\alpha\wedge\beta
\end{equation*}

\subsubsection{Cross norms for Lebesgue spaces}
\label{s:cross-lebesgue}
For a Banach space $(A,\| \cdot \|)$, finite set $I$ and
$p\in[1,\infty]$ denote by
\begin{equation*}
  \ell_{I}^{p}(A) := A^{I}
\end{equation*}
the Banach space of $I$-indexed families of vectors in
$A$ with the norm
\begin{equation*}
  \| (a_{i}) \|_{p,I}:=\left(\sum_{i\in I}\|a_{i}\|^{p}\right)^{1/p}
\end{equation*}
with the usual provision for the case $p=\infty$.
We will write $\ell^{p}_{I}:=\ell^{p}_{I}(\Rbb)$.

An important example for our purposes is
\begin{equation*}
  \ell^{p}_{I}(\ell^{q}_{J})
  =
  \left(\Rbb^{I\times J},
    \|x\|_{p,I;q,J}
    :=
    \Big(\sum_{i\in I}\big(\sum_{j\in J}|x_{ij}|^{q}\big)^{p/q}\Big)^{1/p}\right)
\end{equation*}
with the usual modifications in the cases where $p$ or $q$ is infinite.

It is straightforward to establish, that for finite $p,q$
\begin{equation*}
  \big(\ell^{p}_{I}(\ell^{q}_{J})\big)^{*}\iso[b]\ell^{p'}_{I}(\ell^{q'}_{J})
\end{equation*}
where $p',q'$ are the Hölder conjugates of $p,q$, respectively.
There are canonical algebraic isomorphisms independent of $p,q$
\begin{equation*}
  \ell^{p}_{I}(\ell^{q}_{J})
  \iso[a]
  \ell^{q}_{J}(\ell^{p}_{I})
  \iso[a]
  \Rbb^{I\times J}
\end{equation*}
To avoid ambiguity, we denote the norm on $\ell^{p}_{I}(\ell^{q}_{J})$ by
\begin{equation*}
  \| (x_{ij}) \|_{p,I;q,J}
  :=
  \left(\sum_{i\in I}\left(\sum_{j\in J}|x_{ij}|^{q}\right)^{p/q}\right)^{1/p}
\end{equation*}
Another pair of isomorphisms, that are used in what follows, are
\begin{equation*}
  \ell_{I}^{p}\otimes\ell_{J}^{p}
  \iso[a]
  \ell_{I\times J}^{p}\iso[b]\ell_{I}^{p}(\ell_{J}^{p})
\end{equation*}
We note that, under the algebraic identification above
the Lebesgue $p$-norm on $\ell_{I\times J}^{p}$ is a cross norm with
respect to the tensor product decomposition
$\ell_I^p\otimes \ell_J^p$. This follows directly from the definitions
and duality $(\ell^{p}_{I})^{*}\iso[b]\ell^{p'}_{I}$.

\subsection{Multiplicative family of norms for bilinear forms}
\label{s:mult-family}
Let $I,J$ be finite sets. For a bilinear form
$M:\ell_{I}^{p}\times\ell_{J}^{q}\to\Rbb$ define
\begin{equation*}
  \|M\|_{p,q}
  :=
  \sup\set{\frac{|aMb|}{\|a\|_{p}\cdot\|b\|_{q}}\st
  0\neq a\in \ell_{I}^{p},\;0\neq b\in \ell_{J}^{q}}
\end{equation*}
where we use infix notation for the value $aMb$ of the form $M$ on
vectors $a$, $b$. If we view $M$ as an element from
$\ell_{I}^{p'}\otimes \ell_{J}^{q'}$ or as the operator
$M:\ell_{I}^{p}\to\ell_{J}^{q'}$, then
\begin{equation*}
  \|M\|_{p,q}
  =
  (\|\cdot\|_{p'}\vee\|\cdot\|_{q'})(M)
  =
  \|M\|_{p\to q'}
\end{equation*}

The main technical result, Proposition~\ref{p:multiplicative}, asserts
that in hypercontractive regime, $\frac1p+\frac1q\geq1$, the norms
$\|\cdot\|_{p,q}$ form a multiplicative family for bilinear forms on
$\ell_{I}^{p}\times\ell_{J}^{q}$, where $I,J$ range over finite sets.

\begin{proposition}\label{p:multiplicative}
  Let $p,q\in[1,\infty]$ satisfy $1/p + 1/q \geq 1$ and let $I,J,K,L$
  be finite sets. For two bilinear forms
  \begin{equation*}
    M:\ell^{p}_{I}\times\ell^{q}_{J}\to\Rbb
    \qquad\text{and}\qquad
    N:\ell^{p}_{K}\times\ell^{q}_{L}\to\Rbb
  \end{equation*}
  consider their tensor product
  \[
    M\otimes N:\ell^{p}_{I\times K}\times\ell^{q}_{J\times L}\to\Rbb
  \]
  Then
  \[
    \|M\otimes N\|_{p,q} = \|M\|_{p,q}\cdot\|N\|_{p,q}
  \]
\end{proposition}

To prove the proposition we shall use the following lemma, a
version of the so called Minkowski’s integral inequality,
\cite{hardy1952inequalities}, that says that under canonical algebraic
isomorphism
$\ell^{p}_{I}(\ell^{q}_{J})\iso[a]\ell^{q}_{J}(\ell^{p}_{I})$
norms
$\| \cdot \|_{p,I;q,J}$ and $\| \cdot \|_{q,J;p,I}$ are comparable.
\begin{lemma}\label{p:holder-switch}
  Let $I,J$ be finite sets.
  If $1\leq p\leq q\leq\infty$, then
  \[
    \| \cdot \|_{q,J;p,I}
    \leq
    \| \cdot \|_{p,I;q,J}
  \]
\end{lemma}
We give a proof of the lemma in Section~\ref{s:lemma-proof}.

\subsubsection{Proof of Proposition~\ref{p:multiplicative}}
\label{s:proof-multiplicative}
By renormalization, it is enough to prove that
\begin{equation*}
  \textit{If}\quad
  \|M\|_{p,q}=\|N\|_{p,q}=1,
  \quad\textit{then}\quad
  \|M\otimes N\|_{p,q}=1.
\end{equation*}
First we prove the lower bound. Indeed,
since the Lebesgue norms are cross norms,
\[
  \|a\otimes c\|_{\ell_{I\times K}^p}
  =
  \|a\|_{\ell_I^p}\cdot\|c\|_{\ell_K^p},
  \qquad
  \|b\otimes d\|_{\ell_{J\times L}^q}
  =
  \|b\|_{\ell_J^q}\cdot\|d\|_{\ell_L^q}.
\]
Hence, by testing against decomposable vectors, we have
\[
  \|M\otimes N\|_{p,q}
  \geq
  \|M\|_{p,q}\cdot\|N\|_{p,q}
  =
  1
\]
Now we prove the upper bound.  Let
\[
  x=(x_{ik}\st i\in I,\; k\in K)\in\ell_{I\times K}^p,
  \qquad
  y=(y_{jl}\st j\in J,\; l\in L)\in\ell_{J\times L}^q
\]
be such that $\|x\|_p\leq1$ and $\|y\|_q\leq1$.

For each $k\in K$ and $j\in J$, set
\[
  x_k:=(x_{ik}\st i\in I)\in\ell_I^p,
  \qquad
  y_j:=(y_{jl}\st l\in L)\in\ell_L^q
\]
Since $\|M\|_{p,q}=1$, the associated operator
\[
  M\colon \ell_I^p\to\ell_J^{q'}
\]
has norm $1$.  Hence
\[
  (Mx_k\st k\in K)\in \ell_K^p(\ell_J^{q'})
\]
and
\begin{equation*}
  \|(Mx_k\st k\in K)\|_{p,K;\,q',J}
  \leq
  \|(x_k)_{k\in K}\|_{p,K;\,p,I}
  =
  \|x\|_{\ell_{I\times K}^p}
  \leq1.
\end{equation*}
Since
\[
  \frac1p+\frac1q\geq1
  \qquad\text{iff}\qquad
  p\leq q'
\]
we can apply Lemma~\ref{p:holder-switch}:
\begin{equation}\label{eq:switch-M-fibers-bilinear}
  \|(Mx_k)_{k\in K}\|_{q',J;p,K}
  \leq
  \|(Mx_k)_{k\in K}\|_{p,K;q',J}
  \leq1.
\end{equation}

For each $j\in J$, define
\[
  u_j:=\bigl((Mx_k)_j\bigr)_{k\in K}\in\ell_K^p
\]
Then \eqref{eq:switch-M-fibers-bilinear} says that
\[
  \|(u_j)_{j\in J}\|_{q',J;p,K}\leq1
\]
Therefore, using $\|N\|_{p,q}=1$, we get (with infix notation)
\begin{align*}
  |x(M\otimes N)y|
  &=
    \left|
    \sum_{j\in J}
    u_jNy_j
    \right|
  \\
  &\leq
    \sum_{j\in J}
    \|u_j\|_p\cdot\|y_j\|_q\\
  &\leq
    \|(u_j)_{j\in J}\|_{q',J;p,K}
    \cdot
    \|(y_j)_{j\in J}\|_{q,J;q,L}\\
  &\leq
    1
\end{align*}
Here the last inequality uses Hölder's inequality along $j$-index
and the fact that
\[
  \|(y_j)_{j\in J}\|_{q,J;\,q,L}
  =
  \|y\|_{\ell_{J\times L}^q}
  \leq1.
\]
Thus $\|M\otimes N\|_{p,q}\leq1$, and the proof is complete.
\hfill\qed

\subsubsection{Proof of Lemma~\ref{p:holder-switch}}
\label{s:lemma-proof}
The lemma is immediate when $q=\infty$.  Indeed,
\begin{equation*}
  \|(x_{ij})\|_{\infty,J;p,I}
  =
  \max_{j\in J}\left(\sum_{i\in I}|x_{ij}|^p\right)^{1/p}
  \leq
  \left(\sum_{i\in I}\max_{j\in J}|x_{ij}|^p\right)^{1/p}
  =
  \|(x_{ij})\|_{p,I;\infty,J}.
\end{equation*}
By duality, this also implies the case $p=1$ and any $q$,
that is for any $q$ holds
\begin{equation*}
  \| \cdot \|_{1,I;q,J}\geq\| \cdot \|_{q,J;1,I}
\end{equation*}
For the general case, we shall use the elementary identity
\begin{equation*}\label{eq:mixed-norm-power}
  \|(x_{ij})\|_{p,I;\,q,J}
  =
  \bigl\|(|x_{ij}|^r)\bigr\|_{p/r,I;\,q/r,J}^{1/r},
\end{equation*}
valid for every $0<r\leq \min\set{p,q}$.  In particular, when
$1<p\leq q<\infty$, we may take $r=p$.  Applying the case $p=1$ to
$(|x_{ij}|^p)$ and to the exponent $q/p\geq1$, we obtain
\begin{align*}
  \|(x_{ij})\|_{p,I;\,q,J}
  &=
    \bigl\|(|x_{ij}|^p)\bigr\|_{1,I;\,q/p,J}^{1/p} \\
  &\geq
    \bigl\|(|x_{ij}|^p)\bigr\|_{q/p,J;\,1,I}^{1/p} \\
  &=
    \|(x_{ij})\|_{q,J;\,p,I}.
\end{align*}
This proves the lemma in the general case.
\hfill\qed

\subsection{Resonance vectors for tensor powers of bilinear forms}
\label{s:resonance-tensor}
\subsubsection{Resonance vectors for bilinear form. Uniform vectors}
\label{s:resonance-and-unif}
Consider a bilinear form
$M:\ell^{p}_{\Xsf}\times\ell^{q}_{\Ysf}\to\Rbb$.  We say that a pair
of non-zero vectors $u\in\ell^{p}_{\Xsf}$ and $v\in\ell^{q}_{\Ysf}$ is
a \emph{resonance pair} or equivalently a \emph{maximizing pair} for
the form $M$ if (with infix notation for the value of bilinear forms)
\begin{equation*}
  uMv = \|M\|_{p,q}\cdot \|u\|_{p}\cdot\|v\|_{q}
\end{equation*}
Essentially, a resonance pair consists of the two optimizers in the
definition of the norm of the bilinear form $M$.  Clearly any positive
multiples of the vectors in a resonance pair also form a
resonance pair. In finite dimensions resonance vectors always
exist. If coefficients of $M$ with respect to natural bases are
non-negative then there exist pair of resonance vectors, which also
have non-negative coefficients.

We call a nonzero vector $u\in\ell_{\Xsf}^p$ \emph{uniform}
if it is a positive scalar multiple of the indicator function of a
nonempty subset of $\Xsf$.

We show below that for large tensor powers of a coordinate-wise
non-negative bilinear form $M$, there exists a pair of uniform ``almost''
resonance vectors.  Here adverb ``almost'' should be understood on the
normalized $\log$-scale, see Theorem~\ref{p:almost-resonance-pairs}
below.

\subsubsection{Uniform almost resonance pairs for tensor powers}
\label{s:uniform-resonance}
Throughout this section we fix $p,q\in[1,\infty]$ satisfying
$1/p + 1/q\geq1$, two finite sets $\Xsf$ and $\Ysf$ and a bilinear
form $M:\ell^{p}_{\Xsf}\times\ell^{q}_{\Ysf}\to\Rbb$ of norm one and
with non-negative coefficients with respect to the standard bases in
$\ell^{p}_{\Xsf}$ and $\ell^{q}_{\Ysf}$. We only provide proofs for
finite $p,q$. The results of this section also hold when one of the
exponents $p,q$ is infinite. While proofs in the endpoint cases are
different, they are much simpler and are left to the reader.

% Recall that by Proposition~\ref{p:multiplicative} we have for all $n$
% \[
  % \|M^{\otimes n}\|_{p,q}=\|M\|_{p,q}^{n}=1
% \]

\begin{theorem}\label{p:almost-resonance-pairs}
  Let $p,q\in[1,\infty]$ satisfying $1/p+1/q\geq1$
  and let
  \begin{equation*}
    M:\ell^{p}_{\Xsf}\times\ell^{q}_{\Ysf}\to\Rbb
  \end{equation*}
  be a coordinate-wise non-negative bilinear form of norm one. Then
  there exist sequences of uniform vectors
  $\abar_{n}\in \ell^{p}_{\Xsf^{n}}$ and
  $\bbar_{n}\in\ell^{q}_{\Ysf^{n}}$, $n\in\Nbb$ such that
  \begin{equation*}
    \| \abar_{n}\|_{p}=1,
    \qquad
    \| \bbar_{n}\|_{q}=1,
    \qquad
    \lim_{n\to\infty}\frac1n \log(\abar_{n}M^{\otimes n} \bbar_{n})
    =
    0
  \end{equation*}
\end{theorem}

For an arbitrary coordinate-wise non-negative bilinear form, the theorem
gives the following corollary by rescaling.
\begin{corollary}\label{p:unif-dense}
  Let $p,q\in[1,\infty]$ satisfy $1/p+1/q\geq1$, and let
  $M:\ell^{p}_{\Xsf}\times\ell^{q}_{\Ysf}\to\Rbb$ be a non-zero
  bilinear form with non-negative coefficients with respect to the
  standard bases in $\ell^{p}_{\Xsf}$, $\ell^{q}_{\Ysf}$, where
  $1/p+1/q\geq1$.  Then there exist sequences of uniform vectors
  $\abar_{n}\in \ell^{p}_{\Xsf^{n}}$ and
  $\bbar_{n}\in\ell^{q}_{\Ysf^{n}}$, $n\in\Nbb$ such that
  \begin{equation*}
    \lim_{n\to\infty}\frac1n\Big(\log(\abar_{n}M^{\otimes n}
    \bbar_{n})
    -\log\| \abar_{n}\|_{p}-\log\|
    \bbar_{n}\|_{q}\Big)=\log\|M\|_{p,q}
  \end{equation*}
  By renormalization we can choose $\abar_{n}=\1_{\Asf_{n}}$ and
  $\bbar_{n}=\1_{\Bsf_{n}}$ for suitable subsets $\Asf_{n}\subset\Xsf^{n}$
  and $\Bsf_{n}\subset\Ysf^{n}$. Then
  \begin{equation*}
    \lim_{n\to\infty}\frac1n\Big(\log(\abar_{n}M^{\otimes n}
    \bbar_{n})
    -\frac1p \log|\Asf_{n}|
    - \frac1q\log|\Bsf_{n}|\Big)=\log\|M\|_{p,q}
  \end{equation*}  
\end{corollary}

\begin{remark}\label{r:uniform-optimizers}
  Consider the following extremal problem similar to the definition of
  the norm of a bilinear form
  \[
    \|M^{\otimes n}\|_{p,q}^{(u)}
    :=
    \sup\set{
      \frac{\phi M^{\otimes n}\psi}
      {\|\phi\|_{p}\cdot\|\psi\|_{q}}
      \st
      \phi\in\ell^{p}_{\Xsf^{n}}\text{ and }\psi\in\ell^{p}_{\Ysf^{n}}
      \text{ are uniform}}
  \]
  Clearly we have
  $\|M^{\otimes n}\|_{p,q}^{(u)}\leq\|M^{\otimes n}\|_{p,q}$.  In
  effect, Corollary~\ref{p:unif-dense} states that on the normalized
  $\log$-scale, the usual norm and the restricted subnorm defined
  above are asymptotically the same.

  If necessary we may assume that uniform vectors $\abar_{n}$ and $\bbar_{n}$
  provided by Corollary~\ref{p:unif-dense} are optimizers in the
  definition of the restricted subnorm $\|M^{\otimes n}\|_{p,q}^{(u)}$. 
\end{remark}

\subsubsection{Proof of Theorem~\ref{p:almost-resonance-pairs}}
\label{s:proof-almost-resonance-pairs}
\paragraph{Some recollections.}
We start by recalling  several notions needed in the proof of
Theorem~\ref{p:almost-resonance-pairs}.  For a finite set $\Xsf$
denote by $\Delta\Xsf$ the set of probability distributions on $\Xsf$.
Define the \emph{empirical map}
\[
  \qbf:\Xsf^{n}\to\Delta\Xsf
\]
For $\xbf=(x_{0},\dots,x_{n-1})\in\Xsf^{n}$ the value
of the distribution $\qbf(\xbf)$ at point $x\in\Xsf$ is
\[
  \qbf(\xbf)(x):=\frac{\sharp\set{i\st x_{i}=x}}{n}
\]

Let $\pi\in\Delta\Xsf$ be a distribution on $\Xsf$. Define the
\emph{divergence ball} of radius $r>0$ around $\pi$ as
\begin{equation*}
  B_{r}(\pi)
  :=
  \set{\pi'\st \Dsf(\pi'\sep \pi)\leq r}\subset\Delta\Xsf
\end{equation*}
where $\Dsf$ stands for the Kullback--Leibler divergence. Denote by
$B^{c}_{r}(\pi)$ the complement of the divergence ball. Note that the
support of every distributions in $B_{r}(\pi)$ is contained in the
support of $\pi$.

We shall use the following standard Sanov-type
bound,~\cite{sanov1957large, csiszar1998method}:
\begin{equation}\label{eq:untypical-bound}
  \pi^{\otimes n}(\qbf^{-1}B^{c}_{r}(\pi))
  \leq
  \ebf^{-n\cdot r+|\Xsf|\cdot\log(n+1)}
\end{equation}
For every $\xbf\in\Xsf^n$, the probability of $\xbf$ under the product
distribution $\pi^{\otimes n}$ is
\begin{equation}\label{eq:multi-prob}
  \pi^{\otimes n}(\xbf)
  =
  \ebf^{-n\big[h(\qbf(\xbf))+\Dsf(\qbf(\xbf)\sep \pi)\big]}
\end{equation}
As usual, we use the convention $\ebf^{-\infty}=0$.

\paragraph{The proof.}
We treat the case $p,q<\infty$, leaving the cases of $p$ or $q$
infinite to the reader, as they are much simpler.

Uniform asymptotically resonance vectors $\abar_{n}$ and $\bbar_{n}$
are constructed in three steps.  We start with the ``true''
resonance pair of unit vectors $u$ and $v$ for $M$.  By
Proposition~\ref{p:multiplicative} their tensor powers $u^{\otimes n}$
and $v^{\otimes n}$ form a resonance pair for $M^{\otimes n}$.  In the
first step we restrict $u^{\otimes n}$ and $v^{\otimes n}$ on suitable
typical subsets in $\Xsf^{n}$ and $\Ysf^{n}$, respectively. The
resulting vectors $a_{n}$ and $b_{n}$ are almost unit and
quasi-uniform. In the second step we replace $a_n$ and $b_{n}$ by unit
uniform vectors $\abar_{n}$ and $\bbar_{n}$ with the same respective
supports. In the third step we combine the bounds in the previous
steps to derive the proposition.

\paragraph{Step 1.}  
Let $u\in\ell^{p}_{\Xsf}$ and $v\in\ell^{q}_{\Ysf}$ be unit, coordinate-wise
non-negative, resonance pair for $M$. That is
\begin{align*}
  \text{For all $\xsf\in\Xsf$, $\ysf\in\Ysf$}:
  \quad
  u_{\xsf},v_{\ysf}\geq0,
  \qquad
  \| u \|_{p} = \| v \|_{q} = 1
  \quad\text{and}\quad
  uMv = 1
\end{align*}
We use $u^{\otimes n}$ to construct $a'_{n}\in\ell^{p}_{\Xsf^{n}}$ of
unit norm and with reduced support which forms any asymptotically
resonance pair with $v^{\otimes n}$. The construction of vector
$b'_{n}\in\ell^{q}_{\Ysf^{n}}$ goes along the similar lines.

Since
\begin{equation*}
  \|u\|_{p}^{p} = \sum_{\xsf\in\Xsf}u_{\xsf}^{p}=1
\end{equation*}
we regard $\alpha:=(u_{\xsf}^{p}\st \xsf\in\Xsf)$ as a
probability distribution on $\Xsf$. Consider the sequence of
divergence balls
\[
  B_{\frac1{\sqrt n}}(\alpha)%=B_{n^{-1/2}}(\alpha)
  :=\set{\pi\st \Dsf(\pi\sep\alpha)\leq n^{-1/2}}\subset\Delta\Xsf
\]
of radius $n^{-1/2}$ around $\alpha$, and
denote $B_{\frac1{\sqrt n}}^{c}(\alpha)$ their complements.

Let $D_{n}(\alpha)\subset \Xsf^{n}$ be the preimage of the divergence
ball $B_{\frac1{\sqrt n}}(\alpha)$ under the empirical map
$\qbf:\Xsf^{n}\to\Delta\Xsf$ and let $D^{c}_{n}(\alpha)\subset \Xsf^{n}$
be its complement.  Define
\begin{equation*}
  a_{n}:=u^{\otimes n}|_{D_{n}(\alpha)}
  \qquad\text{and}\qquad
  r_{n}:=u^{\otimes n}|_{D^{c}_{n}(\alpha)}
\end{equation*}
Here $a_{n}$ is the \emph{typical part} and $r_{n}$ is the \emph{atypical remainder}.
The supports of $a_{n}$ and $r_{n}$ are complimentary and we have
\begin{equation*}
  a_{n}+r_{n}=u^{\otimes n}
  \qquad\text{and}\qquad
  \| a_{n}\|_{p}^{p} +\|r_{n}\|_{p}^{p}=\| u^{\otimes n} \|_{p}^{p}=1
\end{equation*}
By Equation~\eqref{eq:untypical-bound} we have
\begin{equation*}
  \|r_{n}\|_{p}^{p}
  =
  \alpha^{\otimes n}\big(D^{c}_{n}(\alpha)\big)
  \leq
  \ebf^{-n^{1/2} + O(\log n)}
  \leq
  \ebf^{-c\cdot n^{1/2}}
\end{equation*}
for some $c>0$ and all sufficiently large $n$.
Thus
\[
  1\geq\|a_{n}\|_{p}\geq 1-\ebf^{-c\cdot n^{1/2}}
\]
Define $a'_{n}:=a_{n}/ \|a_{n}\|_{p}$.

In a similar fashion, write
\[
  \beta:=(v_\ysf^q\st\ysf\in\Ysf)\in\Delta\Ysf
\]
We then similarly decompose $v^{\otimes n}= b_{n}+s_{n}$ into typical
part and atypical remainder. Then
\[
  1\geq\|b_{n}\|_{q}\geq 1-\ebf^{-c\cdot n^{1/2}}
\]
for sufficiently large $n$ and some $c>0$.
Define $b'_{n}:=b_{n}/ \|b_{n}\|_{q}$.

Clearly $a'_{n}$ and $b'_{n}$ are unit vectors in
$\ell^{p}_{\Xsf^{n}}$ and $\ell^{q}_{\Ysf^{n}}$, respectively, and for
sufficiently large $n$ we have
\begin{align}
  \notag
  a'_{n}M^{\otimes n}b'_{n}
  &=
    \|a_{n}\|_{p}^{-1} \|b_{n}\|_{q}^{-1}(a_{n} M^{\otimes n}b_{n})\\
  \label{eq:a'Mb'}
  &=
    \|a_{n}\|_{p}^{-1} \|b_{n}\|_{q}^{-1}
    \Bigl((u^{\otimes n}-r_{n})M^{\otimes n}(v^{\otimes n}-s_{n})\Bigr)
  \\
  \notag
  &=
    \|a_{n}\|_{p}^{-1} \|b_{n}\|_{q}^{-1}
    \Bigl(1-u^{\otimes n}M^{\otimes n}s_{n}-r_{n} M^{\otimes
    n}v^{\otimes n}+r_{n} M^{\otimes n}s_{n}\Bigr)\\
  \notag
  &=1+o(n^{0})
\end{align}

\paragraph{Step 2.}
We will now estimate the quasi-uniformity constant of $a'_{n}$ and
show that it grows subexponentially in $n$. For any
$\xbf\in D_{n}(\alpha)$ Equation~\eqref{eq:multi-prob} gives
\begin{align*}
  a'_{n}(\xbf)
  &=\|a_{n}\|^{-1}_{p}\cdot a_{n}(\xbf)
    =\|a_{n}\|^{-1}_{p}\cdot\big[\alpha^{\otimes n}(\xbf)\big]^{1/p}\\
  &=\|a_{n}\|^{-1}_{p}\cdot
    \ebf^{-\frac1p n[h(\qbf(\xbf))+\Dsf(\qbf(\xbf)\sep\alpha)]}
\end{align*}

First we note that for sufficiently large $n$, the divergence ball
around $\alpha$ of radius $n^{-1/2}$ lies in an open face of the
simplex $\Delta\Xsf$ --- the one that corresponds to the support of
$\alpha$. From now on, we assume that $n$ is sufficiently large for
this assertion to hold. Consequently, there exists a compact subset
$K$ of this open face that contains all divergence balls
$B_{\frac1{\sqrt n}}(\alpha)$ for large $n$. The entropy function
$h:\Delta\Xsf\to\Rbb$ is smooth on $K$. Let $d>0$ be an upper bound
for the norm of differential of $h$ restricted on $K$, where the norm is
evaluated with respect to the total variation distance
$\|\cdot-\cdot\|_{\mathrm{tv}}=\tfrac12\|\cdot-\cdot\|_{1}$ on
$\Delta\Xsf$.
  
Let $\xbf,\xbf'\in D_{n}(\alpha)$. 
Since $\qbf(\xbf),\qbf(\xbf')\in B_{\frac1{\sqrt n}}(\alpha)$, we have
\[
  \Big|\Dsf(\qbf(\xbf)\sep\alpha)-\Dsf(\qbf(\xbf')\sep\alpha)\Big|
  \leq
  n^{-1/2}
\]
By Pinsker inequality,~\cite{pinsker1964information,
  csiszar2011information}, we have
\[
  \|\qbf(\xbf)-\qbf(\xbf')\|_{\mathrm{tv}}\leq \sqrt{2}\cdot n^{-1/4}
\]
and therefore
\[
  \Big|h\big(\qbf(\xbf)\big)-h\big(\qbf(\xbf')\big)\Big|
  \leq
  \sqrt{2}\cdot d\cdot n^{-1/4} 
\]
Thus, we can estimate the logarithm of quasi-uniformity constant for
$a'_{n}$ as follows.
\begin{align*}
  |\log a'_{n}&(\xbf)-\log a'_{n}(\xbf')|
                =
                \frac1p|\log\alpha^{\otimes n}(\xbf)-\log\alpha^{\otimes n}(\xbf')|\\
              &\leq
                \frac1p
                n\Big|h(\qbf(\xbf))-h(\qbf(\xbf'))\Big|
                +
                \frac1p
                n\Big|\Dsf(\qbf(\xbf)\sep\alpha)-\Dsf(\qbf(\xbf')\sep\alpha)\Big|\\
              &\leq
                \frac{\sqrt{2}\cdot d}p \cdot n^{3/4} + \frac1p n^{1/2} 
                \leq
                c\cdot n^{3/4}
\end{align*}
for some positive constant $c$.
Thus
\[
  \frac{a'_{n}(\xbf)}{a'_{n}(\xbf')}\leq \ebf^{c\cdot n^{3/4}}
\]
for all $\xbf,\xbf'\in D_{n}(\alpha)$.  In a similar fashion we
obtain, for all $\ybf,\ybf'\in D_n(\beta)$,
\[
  \frac{b'_{n}(\ybf)}{b'_{n}(\ybf')}\leq \ebf^{c\cdot n^{3/4}}
\]
for some $c>0$.

\paragraph{\bf Step 3.} Let $\abar_{n}$ and $\bbar_{n}$ be uniform
unit vectors in $\ell^{p}_{\Xsf^{n}}$ and $\ell^{p}_{\Ysf^{n}}$, whose
supports coincide with that of $a'_{n}$ and $b'_{n}$,
respectively. Then for all $\xbf\in D_{n}(\alpha)$ and
$\ybf\in D_{n}(\beta)$ holds
  \begin{align*}
    \ebf^{-c\cdot n^{3/4}}
    \leq
    \frac{\abar_{n}(\xbf)}{a'_{n}(\xbf)}
    \leq
    \ebf^{c\cdot n^{3/4}}
    \qquad\text{and}\qquad
    \ebf^{-c\cdot n^{3/4}}
    \leq
    \frac{\bbar_{n}(\ybf)}{b'_{n}(\ybf)}
    \leq
    \ebf^{c\cdot n^{3/4}}
  \end{align*}
  Since coefficients of $M^{\otimes n}$ are non-negative, the above pointwise
  estimates imply
  \begin{equation*}
    \ebf^{-2c\cdot n^{3/4}}
    \leq
    \frac{\abar_{n}M^{\otimes n}\bbar_{n}}
    {a'_{n}M^{\otimes n}b'_{n}}
    \leq
    \ebf^{2c\cdot n^{3/4}}
  \end{equation*}
  Taking normalized logarithm and combining with inequality
  in~\eqref{eq:a'Mb'} we get
  \begin{equation*}
    \lim_{n\to\infty}\frac1n\log(\abar_{n}M^{\otimes n}\bbar_{n})=0
  \end{equation*}
  This finishes the proof of the theorem.
  \hfill\qed

\section{Bilinear form associated to a bipartite graph}
\label{s:graph-bilinear}
In this section we consider the bilinear form associated with a
biregular bipartite graph and explore the relations between uniform
vectors, subgraphs, and extensions of random pairs uniformly supported
on the graph.

Let $\Gsf=(\Xsf\sqcup\Ysf,\Esf)$ be a biregular bipartite
graph. Denote by $M:\Rbb^{\Xsf}\times\Rbb^{\Ysf}\to\Rbb$ the bilinear
form associated with $\Gsf$ defined by
\begin{equation*}
  \phi M\psi
  :=
  \sum_{(\xsf,\ysf)\in\Esf}\phi(\xsf)\psi(\ysf)
\end{equation*}
We refer to $M$ as the \emph{incidence bilinear form}, or simply the
\emph{incidence form}, for the graph $\Gsf$.

\subsection{Norms of incidence bilinear forms}
\label{s:incidence-norms}
Let $\Gsf^{n}:=(\Xsf^{n}\sqcup\Ysf^{n},\Esf^{n})$ stand for the power
of $\Gsf$. The incidence bilinear form of $\Gsf^{n}$ is the tensor
power $M^{\otimes n}$ of the incidence form $M$ of $\Gsf$. By
Proposition~\ref{p:multiplicative} we have the equality
\begin{equation*}
  \log\|M\|_{p,q}=\frac1n\cdot\log\|M^{\otimes n}\|_{p,q}
\end{equation*}
for any $p,q\in[1,\infty]$, satisfying $1/p+1/q\geq1$.

In the complimentary range $1/p+1/q\leq1$, the norms of the
incidence form are completely determined by the sizes of the graph.
\begin{proposition}\label{p:norms-small-pq}
  Let $M$ be the incidence form of a biregular bipartite graph
  $\Gsf=(\Xsf\sqcup\Ysf,\Esf)$. Let $p,q\in[1,\infty]$,
  $1/p+1/q\leq1$.
  Then
  \begin{equation*}
    \|M\|_{p,q}=\frac{|\Esf|}{|\Xsf|^{1/p}\cdot|\Ysf|^{1/q}}
  \end{equation*}
\end{proposition}

Note that the right-hand side in the equality above is multiplicative
with respect to product of bipartite graphs. Thus for incidence
bilinear forms the multiplicativity of the norms,
Proposition~\ref{p:multiplicative}, holds without any restriction on the
exponents $p,q$. This property is specific to incidence forms and does
not hold for general bilinear forms.

\subsubsection{Proof of Proposition~\ref{p:norms-small-pq}}
\label{s:proof-norms-small-pq}
We start by evaluating norms for the extremal values of
exponents
\[
  (p,q)\in\set{(\infty,\infty),(1,\infty),(\infty,1)}
\]
For any $u\in\ell^{\infty}_{\Xsf}$ and
$v\in\ell^{\infty}_{\Ysf}$ we have the inequality
\begin{equation*}
  uMv
  =
  \sum_{(\xsf,\ysf)\in\Esf}u_{\xsf}\cdot v_{\ysf}
  \leq
  |\Esf|\cdot \|u\|_{\infty}\cdot\|v\|_{\infty}
\end{equation*}
with equality for $u=\1_{\Xsf}$, $v=\1_{\Ysf}$. Thus
\begin{equation*}
  \|M\|_{\infty,\infty}=|\Esf|
\end{equation*}
Also
\begin{align*}
  uMv
  =
    \sum_{\xsf\in\Xsf}u_{\xsf}\sum_{\ysf:(\xsf,\ysf)\in\Esf}v_{\ysf}
  \leq
    \|u\|_{1}\cdot d_{1}(\Gsf)\cdot\|v\|_{\infty}
\end{align*}
with equality for $u=\delta_{\xsf_0}$ for some $\xsf_{0}\in\Xsf$, and
$v=\1_{\Ysf}$. Similar inequality holds with the roles of $u$ and $v$
switched.  Thus
\begin{equation*}
  \|M\|_{\infty,\infty}=|\Esf|,\qquad
  \|M\|_{1,\infty}=\frac{|\Esf|}{|\Xsf|},\qquad
  \|M\|_{\infty,1}=\frac{|\Esf|}{|\Ysf|}
\end{equation*}

By Riesz--Thorin interpolation, see~\cite{riesz1927surlesmaxima,
  thorin1939convexity}  or~\cite{bergh1976interpolation} for more
modern treatment, we then have
\begin{equation*}
  \|M\|_{p,q}\leq\frac{|\Esf|}{|\Xsf|^{1/p}\cdot|\Ysf|^{1/q}}
\end{equation*}
On the other hand for $u=\1_{\Xsf}$ and $v=\1_{\Ysf}$ we have
\begin{equation*}
  \|u\|_{p}=|\Xsf|^{1/p},\qquad
  \|v\|_{q}=|\Ysf|^{1/q},\qquad
  uMv=|\Esf|
\end{equation*}
which gives the matching lower bound
\begin{equation*}
  \|M\|_{p,q}\geq\frac{|\Esf|}{|\Xsf|^{1/p}\cdot|\Ysf|^{1/q}}
\end{equation*}
This finishes the proof.
\hfill\qed

\subsection{Uniform almost resonance vectors for incidence forms}
\label{s:unif-resonance}
Recall that Corollary~\ref{p:unif-dense} and
Remark~\ref{r:uniform-optimizers} allow us to find almost resonance
pairs of uniform vectors for large tensor powers of bilinear form $M$,
such that in addition they optimize the restricted supremum in the
definition of $\|M^{\otimes n}\|_{p,q}^{(u)}$. We now apply these
results to the incidence form of a biregular bipartite graph. In that
case, the supports of almost resonance uniform pair of vectors have an
additional regularity property, that we discuss below.

Consider the incidence form $M$ of a biregular bipartite graph
$\Gsf$. Let $a\in\ell^{p}_{\Xsf}$ and $b\in\ell^{q}_{\Ysf}$ be two
uniform vectors attaining the maximum in the definition of
$\|M\|_{p,q}^{(u)}$. Without loss of generality we may assume that $a$
and $b$ are indicator functions of some subsets of $\Xsf$ and $\Ysf$,
respectively.  Denote by
$\Hsf=(\Xsf_{\Hsf}\sqcup\Ysf_{\Hsf},\Esf_{\Hsf})$ the subgraph of
$\Gsf$ induced by the supports $\Xsf_{\Hsf}$ and $\Ysf_{\Hsf}$ of $a$
and $b$, respectively. We choose the supports minimal among
optimizers; then the induced subgraph has no isolated vertices.%
\footnote{The fact that $\Hsf$ has no isolated vertices is automatic
  for $p,q<\infty$ since removing isolated vertices would strictly
  improve the quotient in the definition of restricted subnorm. We
  only need to minimize the support in the case when at least one of
  the exponents is infinite.} %

Then we have the following relations: For any $\xsf\in\Xsf_{\Hsf}$ and
$\ysf\in\Ysf_{\Hsf}$
\begin{align*}
  &\delta_{\xsf}Mb=\deg_{\Hsf}(\xsf),
  &&\|a\|_{p}=|\Xsf_{\Hsf}|^{1/p},
  &&aMb = |\Esf_{\Hsf}|\\
  &aM\delta_{\ysf}=\deg_{\Hsf}(\ysf),
  &&\|b\|_{q}=|\Ysf_{\Hsf}|^{1/q} 
\end{align*}
The quotients
\begin{equation*}
  \frac{|\Esf_{\Hsf}|}{|\Xsf_{\Hsf}|}
  \quad\text{and}\quad
  \frac{|\Esf_{\Hsf}|}{|\Ysf_{\Hsf}|}
\end{equation*}
are equal to the average left and right degree of $\Hsf$,
respectively.

For $\epsilon,\delta\in(0,1]$ we say that $\Gsf$ is
$(\epsilon,\delta)$-quasi-biregular if every vertex in the
left part has degree at least $\epsilon$-fraction of the average left
degree and every vertex in the right part has degree at least
$\delta$-fraction of the average right degree. In other words,
for all $\xsf\in\Xsf$, $\ysf\in\Ysf$ holds
\begin{equation*}
  \deg_{\Gsf}(\xsf)\geq \epsilon\cdot\frac{|\Esf|}{|\Xsf|},
  \qquad
  \deg_{\Gsf}(\ysf)\geq \delta\cdot\frac{|\Esf|}{|\Ysf|},
\end{equation*}

In the next proposition we show that for the uniform
$(p,q)$-optimizing pair of vectors $a$ and $b$ the graph induced by
the supports of the vectors is $(\tfrac1p,\tfrac1q)$-quasi-biregular.
\begin{proposition}\label{p:unif-qregular}
  Let $p,q\in[1,\infty)$ and let $a\in\ell^{p}_{\Xsf}$,
  $a\in\ell^{p}_{\Xsf}$ and $\Hsf\subset\Gsf$ be as above. Then for
  all $\xsf\in\Xsf_{\Hsf}$ and $\ysf\in\Ysf_{\Hsf}$ holds
  \begin{align*}
    \deg_{\Hsf}(\xsf)&\geq\frac{1}{p}\frac{|\Esf_{\Hsf}|}{|\Xsf_{\Hsf}|}\\
    \deg_{\Hsf}(\ysf)&\geq\frac{1}{q}\frac{|\Esf_{\Hsf}|}{|\Ysf_{\Hsf}|}
  \end{align*}
  Thus the degree within $\Hsf$ of any vertex can only deviate down
  from the average degree by a fixed factor and $\Hsf$ is
  $(\frac1p,\frac1q)$-quasi-biregular.
\end{proposition}
Proposition~\ref{p:unif-qregular} is not needed for the proof of the
main result. Nevertheless, we include it here because it gives useful
structural information about uniform optimizers.

\subsubsection{Proof of Proposition~\ref{p:unif-qregular}}
\label{s:proof-unif-qregular}
Let $\xsf\in\Xsf_{\Hsf}$ be an arbitrary vertex in the left part of
$\Hsf$ and $d:=\deg_{\Hsf}(\xsf)$.
Removing the vertex $\xsf$ can not improve the optimization
quotient, hence
\begin{equation*}
  \frac{(a-\delta_{\xsf})Mb}{\|a-\delta_{\xsf}\|_{p}\cdot\|b\|_{q}}
  =
  \frac{|\Esf_{\Hsf}|-d}{(|\Xsf_{\Hsf}|-1)^{1/p}\cdot|\Ysf_{\Hsf}|^{1/q}}
  \leq
  \frac{|\Esf_{\Hsf}|}{|\Xsf_{\Hsf}|^{1/p}\cdot|\Ysf_{\Hsf}|^{1/q}}
  =
   \frac{aMb}{\|a\|_{p}\cdot\|b\|_{q}}
\end{equation*}
Therefore
\begin{equation*}
  \frac{d}{|\Esf_{\Hsf}|}
  \geq
  1-\left(1-\frac1{|\Xsf_{\Hsf}|}\right)^{1/p}
  \geq
  \frac1{p\cdot|\Xsf_{\Hsf}|}
\end{equation*}
which gives
\begin{equation*}
  d\geq\frac1p\cdot\frac{|\Esf_{\Hsf}|}{|\Xsf_{\Hsf}|}
\end{equation*}
Similar argument works for the right part of the subgraph $\Hsf$.
\hfill\qed

\newpage
\subsection{Subgraphs and extensions}
\label{s:subgraphs-ext}
\subsubsection{Construction of an extension from a subgraph}
\label{s:subgraph2extension}
Suppose $\Gsf$ is a homogeneous bipartite graph and $(X,Y)$ is
uniformly supported on $\Gsf$. Denote by $\aut(\Gsf)$ the automorphism
group of $\Gsf$ (acting on $\Gsf$ on the left). Let $\Hsf\subset\Gsf$
be a subgraph without isolated vertices. Applying symmetries
$\sigma\in\aut(\Gsf)$ we obtain a family of subgraphs
\begin{equation*}
  \set{\sigma\Hsf
    :=
    (\sigma\Xsf_{\Hsf}\sqcup\sigma\Ysf_{\Hsf},\sigma\Esf_{\Hsf})\st
  \sigma\in\aut(\Gsf)}
\end{equation*}

We now construct an extension $(X,Y,W)$ with alphabet
$\Wsf:=\aut(\Gsf)$.  The joint distribution $p$ of the triple is defined by
\begin{equation*}
  p(\xsf,\ysf,\sigma)=
  \begin{cases}
    \frac{1}{|\Esf_{\Hsf}|\cdot|\aut(\Gsf)|},
    &\text{if $(\xsf,\ysf)\in\sigma\Esf_{\Hsf}$};\\
    0,
    &\text{otherwise.}
  \end{cases}
\end{equation*}
All the pairs $(X,Y|\sigma)$ for different choices of
$\sigma\in\aut(\Gsf)$ are isomorphic and have distribution
\begin{equation*}
  p(\xsf,\ysf|\sigma)=
    \begin{cases}
    \frac{1}{|\Esf_{\Hsf}|},
    &\text{if $(\xsf,\ysf)\in\sigma\Esf_{\Hsf}$};\\
    0,
    &\text{otherwise}.
  \end{cases}
\end{equation*}
Thus the supporting graph of $(X,Y|\sigma)$ is $\sigma\Hsf$. We also
have
\begin{equation*}
  p(\xsf|\sigma)=\frac{\deg_{\sigma\Hsf}(\xsf)}{|\Esf_{\Hsf}|}
  \qquad\text{and}\qquad
  p(\ysf|\sigma)=\frac{\deg_{\sigma\Hsf}(\ysf)}{|\Esf_{\Hsf}|}
\end{equation*}
Therefore we have the following bounds
\begin{equation}
  \label{eq:subgr-extension}
  \begin{aligned}
    H(X|W)&=H(X|\sigma)\leq[\Xsf_{\Hsf}],
    &&H(XY|W)=H(XY|\sigma)=[\Esf_{\Hsf}],\\
    H(Y|W)&=H(Y\,|\sigma)\leq[\Ysf_{\Hsf}]
  \end{aligned}
\end{equation}
The next proposition establishes the connection between uniform
optimizers and extensions.

\begin{proposition}\label{p:unif-optimizers-extension}
  Let $(X,Y)$ be a pair uniformly supported on a homogeneous bipartite
  graph $\Gsf=(\Xsf\sqcup\Ysf,\Esf)$ with the incidence form $M$. Let
  $p,q\in[0,\infty)$ and $a\in\ell^{p}_{\Xsf}$, $b\in\ell^{q}_{\Ysf}$
  be a pair of indicator vectors which are optimizers in the
  definition of $\|M\|_{p,q}^{(u)}$. Then there exists an
  extension $(X,Y,W)$ such that
  \begin{align*}
    H(X|W)&\leq p\cdot\log\|a\|_{p},
    &&H(XY|W)=\log(aMb),\\
    H(Y|W)&\leq q\cdot\log\|b\,\|_{q}
  \end{align*}
\end{proposition}

\subsubsection{Proof of Proposition~\ref{p:unif-optimizers-extension}}
\label{s:proof-unif-optimizers-extension}
Let $\Hsf=(\Xsf_{\Hsf}\sqcup\Ysf_{\Hsf},\Esf_{\Hsf})$ be the subgraph of
$\Gsf$ induced by the supports $\Xsf_{\Hsf}$ and $\Ysf_{\Hsf}$ of $a$
and $b$, respectively. Then
\begin{equation*}
  p\cdot\log\|a\|_{p}=[\Xsf_{\Hsf}],
  \qquad
  q\cdot\log\|b\|_{p}=[\Ysf_{\Hsf}]
  \quad\text{and}\quad
  \log(aMb)=[\Esf_{\Hsf}]
\end{equation*}
Application of the construction in Section~\ref{s:subgraph2extension}
produces extension $W$, which satisfies the conclusions of the
proposition, by the substitutions above
and~\eqref{eq:subgr-extension}.
\hfill\qed

\section{Proof of the main theorem}
\label{s:main-thm}
%%% Invisible subsection
%%% Without it hyperref complains about subsubsection within section
\refstepcounter{subsection}
\addcontentsline{toc}{subsection}{\protect\numberline{\thesubsection}\mbox{}}
\sectionmark{Invisible}

\begin{theorem}\label{p:difficult}
  Let $(X,Y)$ be a pair of random variables uniformly supported on
  a homogeneous bipartite graph $\Gsf=(\Xsf\sqcup\Ysf,\Esf)$ with the
  incidence form $M$. Let $p,q\in[1,\infty]$. Then
  \begin{equation*}
    \Scal(X,Y)(\alpha,\beta)=\log\|M\|_{p,q}
  \end{equation*}
  where $\alpha=1-1/p$ and $\beta=1-1/q$.
\end{theorem}
The proof of the theorem splits into two cases, the hypercontractive
regime, $1/p+1/q\geq1$, and Hölder regime, $1/p+1/q\leq1$. The proof
strategy in the two cases seems rather different, and we do not know
whether there exists a unified proof covering both.  Most of the
considerations above can be generalized to longer tuples of random
variables and their shapes. For $n$-tuples the hypercontractive
region, $\sum (1-1/p_{i})\leq 1$, and Hölder region,
$\sum 1/p_{i}\leq 1$, are no longer complimentary, if $n>2$. We do not
know, whether generalized theorem holds in the intermediate region of
values of the exponents not included in these two regions.

\subsubsection{Proof of Theorem~\ref{p:difficult}}
\label{s:proof-difficult}
Let $(X,Y)$, $\Gsf=(\Xsf\sqcup\Ysf,\Esf)$ and $M$ be as in the
theorem. Fix $p,q\in[1,\infty]$ and set $\alpha=1-1/p$, $\beta=1-1/q$.
We consider two cases.

\paragraph{Case $\bm{\alpha+\beta\geq 1}$; equivalently, $\bm{1/p + 1/q\leq1}$.}
By~\cite[Proposition 3.4.A]{matveev2026beyond} holds
\begin{equation}\label{eq:shape}
  \begin{aligned}
    \Scal(X,Y)(\alpha,\beta)
    &=
      \alpha\cdot H(X)+\beta\cdot H(Y)- I(X:Y)\\
    &=
      H(XY)-(1-\alpha)\cdot H(X)-(1-\beta)\cdot H(Y)\\
    &=
      [\Esf]-(1-\alpha)[\Xsf]-(1-\beta)[\Ysf]
  \end{aligned}
\end{equation}
where we used identities from Section~\ref{s:graphs} for the last
equality.

On the other hand, by Proposition~\ref{p:norms-small-pq} for
$1/p+1/q\leq1$ we have
\begin{equation}\label{eq:norm}
  \begin{aligned}
    \log\|M\|_{p,q}
    &=
      \log\frac{|\Esf|}{|\Xsf|^{1/p}\cdot|\Ysf|^{1/q}}
    &=
      [\Esf]-\frac1p[\Xsf]-\frac1q[\Ysf]
  \end{aligned}
\end{equation}
Combination of~\eqref{eq:shape} and~\eqref{eq:norm} and substitution
$\alpha=1-1/p$ and $\beta=1-1/q$ gives the conclusion of the theorem
in this case.

\paragraph{Case $\bm{\alpha+\beta<1}$; equivalently, $\bm{1/p + 1/q>1}$.}
We prove two inequalities, which together imply the theorem,
\begin{align}
  \label{eq:SlessM}
  \Scal(X,Y)(\alpha,\beta)&\leq\log\|M\|_{p,q}\\
  \label{eq:SgreaterM}
  \Scal(X,Y)(\alpha,\beta)&\geq\log\|M\|_{p,q}
\end{align}
Let $(X^{n},Y^{n})$ be the pair obtained by taking $n$ independent
copies of $(X,Y)$.
On the one hand by~\cite[Proposition 3.6.A]{matveev2026beyond}
the shape function is additive under independent products, hence
\begin{equation*}
  \Scal(X,Y)(\alpha,\beta)=\frac1n\Scal(X^{n},Y^{n})(\alpha,\beta)
\end{equation*}
On the other hand, Proposition~\ref{p:multiplicative} implies that the
$\log$-norm of $M$ is also stable in the sense that 
\begin{equation*}
  \log\|M\|_{p,q}=\frac1n \log\|M^{\otimes n}\|_{p,q}
\end{equation*}
\paragraph{Proof of inequality~\eqref{eq:SlessM}.}
Suppose that $(X,Y,W)$ is an extension, such that
\begin{equation*}
  H(XY|W)-(1-\alpha) H(X|W)-(1-\beta)H(Y|W)=\Scal(X,Y)(\alpha,\beta)
\end{equation*}
Using stability of the shape function we also have
\begin{equation*}
  H(X^{n}Y^{n}|W^{n})-(1-\alpha) H(X^{n}|W^{n})-(1-\beta)H(Y^{n}|W^{n})
  =
  n\cdot\Scal(X,Y)(\alpha,\beta)
\end{equation*}

By Tropical Asymptotic Equipartition
Property,~\cite{matveev2018asymptotic}, we can replace the extending
variable $W^{n}$ by another extending variable $U_{n}$ such that the
joint distribution of $(X^{n},Y^{n},U_{n})$ is uniform on its support
and the conditional entropy profiles $e(X^{n},Y^{n}|W^{n})$ and
$e(X^{n},Y^{n}|U_{n})$ are close on the normalized scale.  More
precisely, for every $\epsilon>0$ there exists $n$ and an extension
$(X^{n},Y^{n},U_{n})$, uniform on its support, such that
\begin{align*}
  \left|H(XY|W)-\frac1n H(X^{n}Y^{n}|U_{n})\right|
  &\leq
    \epsilon\\
  \left|H(X|W)-\frac1n H(X^{n}|U_{n})\right|
  &\leq
    \epsilon\\
  \left|H(Y|W)-\frac1n H(Y^{n}|U_{n})\right|
  &\leq
    \epsilon
\end{align*}
Consequently,
\begin{equation*}
  \frac1n\Big(
  H(X^{n}Y^{n}|U_{n})-(1-\alpha)H(X^{n}|U_{n})-(1-\beta)H(Y^{n}|U_{n})\Big)
  \geq \Scal(X,Y)(\alpha,\beta)-3\epsilon 
\end{equation*}
Choose an arbitrary atom $\usf$ from the alphabet of $U_{n}$ and let
$\Hsf\subset\Gsf^{n}$ be the subgraph supporting
$(X^{n},Y^{n}|\usf)$. Write $\Hsf=:(\Lsf\sqcup\Rsf,\Fsf)$, where
\begin{equation*}
  \Lsf\subset\Xsf^{n},\qquad
  \Rsf\subset\Ysf^{n},\qquad
  \Fsf\subset\Esf^{n}
\end{equation*}
are the left part, the right part and the edge set of $\Hsf$. Since
$(X^{n},Y^{n},U_{n})$ is uniform, we have the following identities
\begin{align*}
  H(X^{n}|U_{n})=H(X^{n}|\usf)=[\Lsf]\\
  H(Y^{n}|U_{n})=H(Y^{n}|\usf)=[\Rsf]\\
  H(X^{n}Y^{n}|U_{n})=H(X^{n}Y^{n}|\usf)=[\Fsf]  
\end{align*}
Let $\phi\in\ell^{p}_{\Xsf^{n}}$ and $\psi\in\ell^{q}_{\Ysf^{n}}$ be
the indicator functions of $\Lsf$ and $\Rsf$, respectively.  Then
\begin{equation*}
  \log\|\phi\|_{p}=\frac1p[\Lsf],
  \qquad
  \log\|\psi\|_{q}=\frac1q[\Rsf],
  \qquad
  \log(\phi M^{\otimes n}\psi)\geq[\Fsf]
\end{equation*}
Therefore
\begin{align*}
  \log\|M\|_{p,q}
  &=
    \frac1n\log\|M^{\otimes n}\|_{p,q}\\
  &\geq
    \frac1n
    \Big(
    \log(\phi M^{\otimes n}\psi) - \log\|\phi\|_{p}-\log\|\psi\|_{q}
    \Big)\\
  &\geq
    \frac1n
    \Big(
    [\Fsf] -
    \frac1p[\Lsf]-\frac1q[\Rsf]
    \Big)\\
  &=
    \frac1n
    \Big(
    H(X^{n}Y^{n}|U_{n}) -
    (1-\alpha)H(X^{n}|U_{n})-(1-\beta)H(Y^{n}|U_{n})
    \Big)\\
  &\geq
    \Scal(X,Y)(\alpha,\beta)-3\epsilon
\end{align*}
This inequality holds for every $\epsilon>0$, so
inequality~\eqref{eq:SlessM} follows.

\paragraph{Proof of inequality~\eqref{eq:SgreaterM}.}
By Corollary~\ref{p:unif-dense} for every $\epsilon>0$ there exist
$n\in\Nbb$ and indicator functions $\abar\in\ell^{p}_{\Xsf^{n}}$ and
$\bbar\in\ell^{q}_{\Ysf^{n}}$ such that
\begin{equation*}
  \log\|M\|_{p,q}
  \leq
  \frac1n\Big(\log(\abar M^{\otimes n}\bbar) -
  \log\|\abar\|_{p}-\log\|\bbar\|_{q}\Big)+\epsilon
\end{equation*}
By Proposition~\ref{p:unif-optimizers-extension} there exists an
extension $(X^{n},Y^{n},W)$ such that
\begin{align*}
  &H(X^{n}Y^{n}|W)=\log(\abar M^{\otimes n}\bbar)\\
  &H(X^{n}|W)\leq p\cdot\log\|\abar\|_{p}\\
  &H(Y^{n}|W)\,\leq q\cdot\log\|\bbar\|_{q}
\end{align*}
Now we can estimate
\begin{align*}
  \log\|M\|_{p,q}
  &\leq\frac1n
  \Big(
    \log(\abar M^{\otimes n}\bbar) -
    \log\|\abar\|_{p}-\log\|\bbar\|_{q}
    \Big)
    +\epsilon\\
  &\leq
    \frac1n
    \Big(
    H(X^{n}Y^{n}|W) -
    \frac1p H(X^{n}|W)- \frac1q H(Y^{n}|W)
    \Big)
    +\epsilon\\
  &\leq
    \frac1n\Scal(X^{n},Y^{n})(\alpha,\beta)+\epsilon\\
  &=
    \Scal(X,Y)(\alpha,\beta)+\epsilon
\end{align*}
Since $\epsilon>0$ is arbitrary, inequality~\eqref{eq:SgreaterM}
follows.
\hfill\qed

\section{Properties of the shape}
\label{s:properties}
In this section we establish inequalities and relations satisfied by
shapes of pairs of random variables. Most of these relations are
elementary and can be derived directly in the context of the
definition of the shape by using properties of entropy and Shannon
inequalities.  However, inequality~\eqref{eq:comp} and its
implication, inequality~\eqref{eq:ref}, seem to be different. We do
not know, whether a proof not referring to Theorem~\ref{p:difficult}
exists.

Proofs of all statements in Section~\ref{s:proplist} are given in
Section~\ref{s:propproofs}.

\subsection{Lower and upper bounds for the shape function}
We need to recall some definitions from~\cite{matveev2026beyond}: For
a pair of random variables $(X,Y)$ and $(\alpha,\beta)\in[0,1]^{2}$
define
\begin{align*}
  &\Scal_{\min}(X,Y)(\alpha,\beta)
    :=
    \max
    \left\{
    \begin{aligned}
      &\alpha\cdot H(X|Y),\\
      &\beta\cdot H(Y|Y),\\
      &\alpha\cdot H(X)+\beta\cdot H(Y)-I(X:Y)
    \end{aligned}
        \right\}\\
  &\Scal_{\max}(X,Y)(\alpha,\beta)
    :=
    \max
    \left\{
    \begin{aligned}
      &\alpha\cdot H(X|Y)+
        \beta\cdot H(Y|Y),\\
      &\alpha\cdot H(X)+\beta\cdot H(Y)-I(X:Y)
    \end{aligned}
        \right\}  \\[1ex]
  &\ing(X,Y,A,B):=I(X:Y|A)+I(X:Y|B)+I(A:B)-I(X:Y)
\end{align*}
The apex point where three affine pieces of $\Scal_{\min}$ meet is
\begin{align*}
  &\alpha_{0}=\frac{H(Y|X)\cdot I(X:Y)}{H(X)\cdot H(Y)-I(X:Y)^{2}},
  &&\beta_{0}=\frac{H(X|Y)\cdot I(X:Y)}{H(X)\cdot H(Y)-I(X:Y)^{2}},
  \\[1ex]
  &\Scal_{\min}(X,Y)(\alpha_{0},\beta_{0})
    =
    \frac{H(X|Y)\cdot H(Y|X)\cdot I(X:Y)}{H(X)\cdot H(Y)-I(X:Y)^{2}}
    \mkern-200mu
\end{align*}

\subsection{Properties of the shape}
\label{s:proplist}
For every tuple of random variables $(X,Y,X',Y',X'',Y'',Z,A,B)$ and every\\
$t,\alpha,\beta,\alpha',\beta',\alpha'',\beta'',\gamma,\eta\in[0,1]$
holds:

\subsubsection{Symmetry}
\begin{equation}
  \tag{\textbf{SYM}}\label{eq:sym}
  \Scal(X,Y)(\alpha,\beta)=\Scal(Y,X)(\beta,\alpha)
\end{equation}

\subsubsection{Convexity}
If
\[
  \vect{\alpha''\\\beta''}=t\vect{\alpha\\\beta}+(1-t)\vect{\alpha'\\\beta'}
\]
then
\begin{equation}\tag{\textbf{CONV}}\label{eq:conv}
  \Scal(X,Y)(\alpha'',\beta'')
  \leq
  t\cdot\Scal(X,Y)(\alpha,\beta)
  +(1-t)\cdot\Scal(X,Y)(\alpha',\beta')
\end{equation}

\subsubsection{Coordinate-wise monotonicity and reverse monotonicity}
If $\alpha\leq\alpha'$ and $\beta\leq\beta'$ then

\begin{align}\tag{\textbf{MO}}\label{eq:mo}
  &\Scal(X,Y)(\alpha,\beta)
    \leq
    \Scal(X,Y)(\alpha',\beta')
  \\[0.5em]
  \tag{\textbf{RMO}}\label{eq:rmo}
  &\Scal(X,Y)(\alpha',\beta')
    \leq
    \Scal(X,Y)(\alpha,\beta)
    +(\alpha'-\alpha)H(X)+(\beta'-\beta)H(Y)
\end{align}
Written together
\begin{equation*}%\tag{\textbf{RMO}}\label{eq:mo*}
  0
  \leq
  \Scal(X,Y)(\alpha',\beta')-\Scal(X,Y)(\alpha,\beta)
  \leq
  (\alpha'-\alpha)H(X)+(\beta'-\beta)H(Y)
\end{equation*}

\subsubsection{Additivity}
If $(X'',Y'')=(X,Y)\oplus(X',Y')$ is the independent sum, then
\begin{equation}\tag{\textbf{ADD}}\label{eq:add}
  \Scal(X'',Y'')(\alpha,\beta)=\Scal(X,Y)(\alpha,\beta)+\Scal(X',Y')(\alpha,\beta)
\end{equation}

\subsubsection{Chain rule}
\begin{equation}\tag{\textbf{CHAIN}}\label{eq:chain}
  \Scal(XZ,YZ)(\alpha,\beta)=\Scal(X,Y|Z)(\alpha,\beta)+\Scal(Z,Z)(\alpha,\beta)
\end{equation}
% Recall
% \[
  % \Scal(Z,Z)(\alpha,\beta)=\max\set{0,(\alpha+\beta-1)H(Z)}
% \]

\subsubsection{Test inequality}
\begin{equation}\tag{\textbf{TEST}}\label{eq:test}
  \alpha\cdot H(X|Z)+\beta\cdot H(Y|Z)-I(X:Y|Z)
  \leq
  \Scal(X,Y)(\alpha,\beta)
\end{equation}

\subsubsection{Upper bound}
\begin{equation}\tag{\textbf{UP}}\label{eq:up}
  \Scal(X,Y)(\alpha,\beta)
  \leq
  \Scal_{\max}(X,Y)(\alpha,\beta)
\end{equation}
The inequality is equality on the upper-right triangle
$\set{0\leq\alpha,\beta\leq1,\;\alpha+\beta\geq1}$, where the second
max-summand dominates. It is also equality on the boundary of the
square where $\Scal$ is equal to the first max-summand.

The upper bound is attained if and only if mutual information of
$(X,Y)$ is extractable, that is there is an extension
$(X,Y,W)$ with
\[
  H(W|X)=H(W|Y)=I(X:Y|W)=0
\]

\subsubsection{Lower bound}
\begin{equation}\tag{\textbf{LO}}\label{eq:lo}
  \Scal_{\min}(X,Y)(\alpha,\beta)
  \leq
  \Scal(X,Y)(\alpha,\beta)
\end{equation}
On the upper-right triangle and the boundary of the square we have
equality.

Lower bound is attained for rigid pairs. It is attained up to
$O\bigl(\log H(XY)\bigr)$ for pairs supported on balanced expanders,
see~\cite{matveev2026beyond}.

Define the rigidity of the nondegenerate pair $(X,Y)$ by
\begin{equation*}
  \Rcal(X,Y)
  :=
  \frac{\Scal_{\max}(X,Y)(\alpha_{0},\beta_{0})-\Scal(X,Y)(\alpha_{0},\beta_{0})}
  {\Scal_{\max}(X,Y)(\alpha_{0},\beta_{0})-\Scal_{\min}(X,Y)(\alpha_{0},\beta_{0})}
\end{equation*}
Thus we have $0\leq \Rcal(X,Y)\leq 1$, with $\Rcal(X,Y)= 1$ if and
only if the pair is rigid and $\Rcal(X,Y)= 0$ if and only if the pair
has extractable mutual information.

\subsubsection{Composition inequality}
\begin{equation}\tag{\textbf{COMP}}\label{eq:comp}
  \Scal(X,Z)(\alpha,\gamma)
  \leq
  \Scal(X,Y)(\alpha,\beta)+\Scal(Y,Z)(1-\beta,\gamma)-H(Y|XZ)
\end{equation}
In the Hölder regime ($\alpha+\gamma\geq1$), take $\beta$ to be any
value in the interval $[1-\alpha,\gamma]$. The difference between the
right-hand side and the left-hand side is then $I(X:Z|Y)$.

\subsubsection{Refinement and coarsening}
For a pair of random variables $(A,B)$ write $A\preceq B$ if $B$ is a
refinement of $A$, equivalently $H(A|B)=0$.
\medskip

If $X\preceq X'$ and $Y\preceq Y'$ then
\begin{equation}
  \tag{\textbf{REF}}\label{eq:ref}
  \Scal(X,Y)(\alpha,\beta)
  \leq
  \Scal(X',Y')(\alpha,\beta)
  +(1-\alpha)H(X'|X)+(1-\beta)H(Y'|Y) - H(X'Y'|XY)
\end{equation}

\begin{equation}
  \tag{\textbf{CRS}}\label{eq:crs}
  \Scal(X,Y')(\alpha,\beta)
  \leq
  \Scal(X,Y)(\alpha,\beta)
  + \beta\cdot H(Y'|XY)
\end{equation}

We can use symmetry to sequentially coarsen the first and the second variable, but
the resulting bound depend on the order of application. However the
following symmetric but weaker inequality holds
\begin{equation}
  \tag{\textbf{CRS2}}\label{eq:crs2}
  \Scal(X',Y')(\alpha,\beta)
  \leq
  \Scal(X,Y)(\alpha,\beta)
  +\alpha\cdot H(X'|XY)+\beta\cdot H(Y'|XY)
\end{equation}

\subsubsection{Link and reversed link inequalities}
\begin{align}
  \tag{\textbf{LI}}\label{eq:li}
  \Scal(X,Y)(\alpha,\beta)
  &\leq
  \Scal(X,Y|Z)(\alpha,\beta) + I(XY:Z)
  \\[0.5em]
  \tag{\textbf{RLI}}\label{eq:rli}
  \Scal(X,Y|Z)(\alpha,\beta)
  &\leq
  \Scal(X,Y)(\alpha,\beta)
\end{align}

\subsubsection{Flattening/links and reverse}
\begin{align}
  \tag{\textbf{FLI}}\label{eq:fli}
  \Scal(X,YZ)(\alpha,\beta)
  &\leq
    \Scal(X,Y|Z)(\alpha,\beta)+\beta\cdot H(Z)
  \\[0.5em]
  \tag{\textbf{RFLI}}\label{eq:rfli}
  \Scal(X,Y|Z)(\alpha,\beta)
  &\leq
  \Scal(X,YZ)(\alpha,\beta)
\end{align}

\subsubsection{MMRV-inequality}
\begin{equation}
  \tag{\textbf{MMRV}}\label{eq:mmrv}
  3\Scal(X,Y)(\tfrac13,\tfrac13)-2\Scal(X,Y)(\tfrac12,\tfrac12)
  \leq
  \ing(X,Y,A,B)
\end{equation}

\subsection{Proofs}
\label{s:propproofs}
Symmetry~\eqref{eq:sym} follows directly from the definition of the shape.
Convexity of the shape function, \eqref{eq:conv}, is proven in~\cite[Section
2.2]{matveev2026beyond}. The relations \eqref{eq:add} and
\eqref{eq:chain} are proven in~\cite[Section
3.6]{matveev2026beyond}. The upper and lower bounds for the shape
function, \eqref{eq:up} and \eqref{eq:lo}, are proven
in~\cite[Sections 3.4 and 3.5]{matveev2026beyond}.
Inequality~\eqref{eq:test} follows directly
from the definition.

\paragraph{Proof of \eqref{eq:mo} and \eqref{eq:rmo}.} For $\alpha\leq\alpha'$
and $\beta\leq\beta'$ and all $W$'s extending $(X,Y)$ holds
\begin{align*}
  \alpha'\cdot H(X|W) - (\alpha'-\alpha) H(X)
  &\leq
    \alpha\cdot H(X|W)
    \leq
    \alpha'\cdot H(X|W)\\
  \beta'\cdot H(Y|W) - (\beta'-\beta) H(Y)
  &\leq
    \beta\cdot H(Y|W)\leq \beta'\cdot H(Y|W)
\end{align*}
Substituting in the definition of the shape we obtain the required inequalities.
\hfill\qed

\paragraph{Proof of \eqref{eq:comp}.}
By additivity and continuity of the shape function,~\cite[Proposition
3.6.A]{matveev2026beyond} and by Tropical Asymptotic Equipartition
Property,~\cite[Theorem 6.1]{matveev2018asymptotic} we can assume that
the triple $(X,Y,Z)$ is uniform on the support. Let
$p,q,r\in[1,\infty]$ and
\begin{equation*}
  M_{XY}:\ell^{p}_{\Xsf}\to\ell^{q}_{\Ysf},
  \quad
  M_{Y\!Z}:\ell^{q}_{\Ysf}\to\ell^{r}_{\Zsf},
  \quad\text{and}\quad
  M_{X\!Z}:\ell^{p}_{\Xsf}\to\ell^{r}_{\Zsf}
\end{equation*}
be the incidence operators of the graphs supporting pairs $(X,Y)$,
$(Y,Z)$ and $(X,Z)$, respectively.  Denote by $N=M_{Y\!Z}\circ M_{XY}$
the composition of the operators. Then
\begin{equation*}
  N_{\xsf,\zsf}
  =
  \sum_{\ysf\in\Ysf}(M_{Y\!Z})_{\ysf,\zsf}\cdot(M_{XY})_{\xsf,\ysf}
  \geq
  \sharp(\Zsf|\Xsf\Ysf)\cdot(M_{X\!Z})_{\xsf,\zsf}
\end{equation*}

We note that all operators have non-negative coefficients, therefore
coordinate-wise domination implies the corresponding inequality for the
norms.
Thus, taking the operator norms, we obtain inequality 
\begin{equation*}
  \|M_{XY}\|_{p\to q}\cdot\|M_{Y\!Z}\|_{q\to r}
  \geq
  \|N\|_{p\to r}
  \geq
  \sharp(\Zsf|\Xsf\Ysf)\cdot\|M_{X\!Z}\|_{p\to r}
\end{equation*}
Switching to norms of bilinear forms we obtain
\begin{equation*}
  \|M_{XY}\|_{p,q'}\cdot\|M_{Y\!Z}\|_{q,r'}\geq\sharp(\Zsf|\Xsf\Ysf)\cdot\|M_{X\!Z}\|_{p,r'}
\end{equation*}
We now take logarithm of the last inequality, use
Theorem~\ref{p:difficult} and the substitutions $\alpha = 1-1/p$, $\beta
= 1/q = 1-1/q'$ and $\gamma = 1/r = 1-1/r'$, to obtain inequality~\eqref{eq:comp}.
\hfill\qed

\paragraph{Proof of \eqref{eq:ref}.}
We apply~\eqref{eq:comp} to the triple $(X,Y',Y)$. In the calculation below the
summand $\Scal(Y',Y)(1-\beta,\beta)$ is in Hölder regime and its value
is forced by Shannon inequalities.
\begin{align*}
  \Scal(X,Y)(\alpha,\beta)
  &\leq
    \Scal(X,Y')(\alpha,\beta)+\Scal(Y',Y)(1-\beta,\beta)-H(Y'|XY)\\
  &=
    \Scal(X,Y')(\alpha,\beta)
    +(1-\beta)H(Y')+\beta\cdot H(Y)-I(Y':Y)
    -H(Y'|XY)\\
  &=
    \Scal(X,Y')(\alpha,\beta)
    +(1-\beta)H(Y'|Y)
    -H(Y'|XY)
\end{align*}
Applying the above bound twice we get
\begin{align*}
  \Scal(X,Y)&(\alpha,\beta)
  \leq
    \Scal(X,Y')(\alpha,\beta)
    +(1-\beta)H(Y'|Y)
    -H(Y'|XY)\\
  &\leq
    \Scal(X',Y')(\alpha,\beta)
    +(1-\alpha)H(X'|X)+(1-\beta)H(Y'|Y)
    -H(Y'|XY) - H(X'|XY')\\
  &=
    \Scal(X',Y')(\alpha,\beta)
    +(1-\alpha)H(X'|X)+(1-\beta)H(Y'|Y)
    -H(X'Y'|XY)
\end{align*}
This finishes the proof of~\eqref{eq:ref}.
\hfill\qed

\paragraph{Proof of \eqref{eq:crs}.}
Let $W_{0}$ be an optimizer in the definition of
$S(X,Y')(\alpha,\beta)$. Then
\begin{align*}
  \Scal(X,Y')(\alpha,\beta)
  &=
    \alpha\cdot H(X|W_{0})+\beta\cdot H(Y'|W_{0})-I(X:Y'|W_{0})\\
  &=
    \alpha\cdot H(X|W_{0})+\beta\cdot H(Y|W_{0})-I(X:Y|W_{0})\\
  &\quad
    +\beta\big(H(Y'|W_{0})-H(Y|W_{0})\big)
    -\big(I(X:Y'|W_{0})-I(X:Y|W_{0})\big)\\
  &\leq
    \Scal(X,Y)(\alpha,\beta)+\beta\cdot H(Y'|XY)
\end{align*}
\hfill\qed

\paragraph{Proof of \eqref{eq:li}.}
Let $W_{0}$ be an optimizer in the definition of
$\Scal(X,Y)(\alpha,\beta)$, that is
\begin{equation*}
  \Scal(X,Y)(\alpha,\beta)
  =
  H(XY|W_{0})- (1-\alpha)H(X|W_{0})-(1-\beta)H(Y|W_{0})
\end{equation*}
By Mat\'u\v s' inner adhesivity property of entropic polymatroids,
\cite{matuvs2005inequalities,matus2007infinitely}, we can assume that
$W_{0}$ is chosen so that $I(W_{0}:Z|XY)=0$. In that case holds
\begin{equation*}
  I(XY:Z|W_0)=I(XY:Z)-I(W_0:Z)\leq I(XY:Z).
\end{equation*}
Now we can estimate
\begin{align*}
  \Scal(X,Y)(\alpha,\beta)
  &=
    H(XY|W_{0})- (1-\alpha)H(X|W_{0})-(1-\beta)H(Y|W_{0})\\
  &=
    H(XY|ZW_{0})- (1-\alpha)H(X|ZW_{0})-(1-\beta)H(Y|ZW_{0})\\
  &\quad
    +I(XY:Z|W_{0})-(1-\alpha)I(X:Z|W_{0})-(1-\beta)I(Y:Z|W_{0})\\
  &\leq
    \Scal(X,Y|Z)(\alpha,\beta)+I(XY:Z|W_{0})\\
  &\leq
    \Scal(X,Y|Z)(\alpha,\beta)+I(XY:Z)
\end{align*}
which is the required inequality.
\hfill\qed

\paragraph{Proof of \eqref{eq:rli}.}
Since inequality~\eqref{eq:test} holds for any choice of $Z$ extending
$(X,Y)$, we can formally replace $Z$ by $ZW$ to obtain
\begin{equation*}
  \alpha\cdot H(X|WZ) + \beta\cdot H(Y|WZ) - \alpha\cdot I(X:Y|WZ)
  \leq
  S(X,Y)(\alpha,\beta)
\end{equation*}
Taking supremum with respect to $W$ we obtain~\eqref{eq:rli}.
\hfill\qed

\paragraph{Proof of \eqref{eq:fli}.}
Suppose $W_{0}$ is an optimizer in the definition of
$\Scal(X,Y\!Z)(\alpha,\beta)$. Then
\begin{align*}
  \Scal(X,Y\!Z)(\alpha,\beta)
  &=
    \alpha\cdot H(X|W_{0})+\beta\cdot H(Y\!Z|W_{0})-I(X:Y\!Z|W_{0})\\
  &=
    \alpha\cdot H(X|ZW_{0})+\beta\cdot H(Y|ZW_{0})-I(X:Y|ZW_{0})\\
  &\quad
    +\alpha\cdot I(X:Z|W_{0}) + \beta\cdot H(Z|W_{0}) - I(X:Z|W_{0})\\
  &\leq
    \alpha\cdot H(X|ZW_{0})+\beta\cdot H(Y|ZW_{0})-I(X:Y|ZW_{0})\\
  &\quad
    +\beta\cdot H(Z)\\
  &\leq
    \Scal(X,Y|Z)(\alpha,\beta)+\beta\cdot H(Z)
\end{align*}
\hfill\qed

\paragraph{Proof of \eqref{eq:rfli}.}
Let $W_{0}$ be the optimizer in the definition of $\Scal(X,Y|Z)(\alpha,\beta)$. Then
\begin{align*}
  \Scal(X,Y|Z)(\alpha,\beta)
  &=
    H(XY|ZW_{0})-(1-\alpha)H(X|ZW_{0})-(1-\beta)H(Y|ZW_{0})\\
  &=
    H(XY\!Z|ZW_{0})-(1-\alpha)H(X|ZW_{0})-(1-\beta)H(Y\!Z|ZW_{0})\\
  &\leq
    \Scal(X,Y\!Z)(\alpha,\beta)
\end{align*}
\hfill\qed

\paragraph{Proof of \eqref{eq:mmrv}.}
In~\cite{makarychev2002new} the following non-Shannon inequality for
five random variables is proven
\begin{equation*}
  \ing(X,Y,A,B)\geq -I(X:Y|W)-I(W:Y|X)-I(X:W|Y)
\end{equation*}
The right-hand side can be rewritten as
\begin{equation*}
  \begin{aligned}
    -I(X:Y|W)&-I(W:Y|X)-I(X:W|Y)\\
    &= 
  \big(H(X|W ) + H(Y|W ) - 3I(X:Y|W)\big)\\
    &\quad
      - \big(H(X|Y) + H(Y|X)\big)
  \end{aligned}
\end{equation*}
Thus
\begin{equation*}
  \ing(X,Y,A,B)\geq \big(H(X|W ) + H(Y|W ) - 3I(X:Y|W)\big)
  - \big(H(X|Y) + H(Y|X)\big)
\end{equation*}
Since $\Scal(X,Y)(\tfrac12,\tfrac12)=\frac12\big(H(X|Y) + H(Y|X)\big)$, then
taking the supremum over $W$'s gives
\begin{equation*}
  \ing(X,Y,A,B)\geq 3\Scal(X,Y)(\tfrac13,\tfrac13)-2\Scal(X,Y)(\tfrac12,\tfrac12)
\end{equation*}
\hfill\qed

\printbibliography[heading=bibintoc]
\end{document}